\documentclass[usenatbib]{mn2e}
\usepackage{color}
\usepackage{natbib}
\usepackage{paralist}
\usepackage{graphicx}
\usepackage{epstopdf}
\usepackage{epsfig}
\usepackage{amsmath}
\usepackage{tablefootnote}

\def\ciso{{$^{12}$C/$^{13}$C}}

\def\teff{\mbox{$T_{\rm eff}$}}
\def\logg{\mbox{log~{\it g}}}
\def\vmicro{\mbox{$\xi_{\rm t}$}}
\def\kmsec{\mbox{km~s$^{\rm -1}$}}

\def\eg{\mbox{e.g.}}

\title[Chemical compositions of RGs in the 
NGC 6940]{THE CHEMICAL COMPOSITIONS AND EVOLUTIONARY STATUS OF RED GIANTS 
IN THE OPEN CLUSTER NGC 6940}
\author[G. B\"{o}cek Topcu, M. Af\c{s}ar,  C. Sneden]
{G. B\"{o}cek Topcu$^{1}$\thanks{E-mail: gamzebocek@gmail.com (GBT); 
melike.afsar@ege. edu.tr (MA); chris@verdi.as.utexas.edu (CS)}, 
M. Af\c{s}ar$^{1}$, , C. Sneden$^{2}$\\ 
$^{1}$Department of Astronomy and Space Sciences, 
                 Ege University, 35100 Bornova, \.{I}zmir, Turkey;\\
$^{2}$Department of Astronomy and McDonald Observatory,
                 The University of Texas, Austin, TX 78712}
\begin{document}

\date{Published 08 August 2016}

\maketitle

\label{firstpage}

\begin{abstract}

We present the high resolution 
(R $\approx$ 60 000), high 
signal-to-noise (S/N $\simeq$ 120) spectroscopic analysis of 12 red 
giant members of the Galactic open cluster NGC~6940. 
We applied Yonsei-Yale isochrones to the colour-magnitude diagram, which 
suggested an age of 1.1 Gyr for the cluster with a 
turn-off mass of 2~M$_{\odot}$.
Atmospheric parameters (\teff, \logg, [Fe/H] and $\xi_{t}$) were determined 
via equivalent widths of \mbox{Fe\,{\sc i}}, \mbox{Fe\,{\sc ii}}, 
\mbox{Ti\,{\sc i}}, and \mbox{Ti\,{\sc ii}} lines. 
Calculated mean metallicity of the cluster is 
$\langle {\rm [Fe/H]} \rangle = 0.04\pm0.02$.
We derived abundances of $\alpha$ (Mg, Si, Ca), Fe-group (Sc, Ti, V, Cr, Mn, Fe, Co, Ni, Cu, Zn), 
and $n$-capture (Y, La, Nd, Eu) elements to be about solar. 
Light odd-Z elements Na and Al are slightly enhanced in MMU~108 
and MMU~152 by $\sim$0.34 dex and $\sim$0.16 dex, respectively.
Abundances of light elements Li, C, N, O, and \ciso\ ratios were derived from 
spectrum syntheses of the \mbox{Li\,{\sc i}} resonance doublet at 6707 \AA, 
[{\mbox{O\,{\sc i}}}] line at 6300 \AA, C$_{2}$ Swan bandheads at 5164 \AA\ 
and 5635 \AA, and strong $^{12,13}$CN system lines 
in the 7995$-$8040 \AA\ region. 
Most carbon isotopic ratios are similar to those found in other 
solar-metallicity giants, but MMU~152 has an unusual value of \ciso~$=6$. 
Evaluation of the LiCNO abundances and \ciso\ ratios along with the present 
theoretical models suggests that all the red giants in our sample are 
core-helium-burning clump stars.

\end{abstract}

\begin{keywords}
stars: abundances -- stars:  atmospheres. Galaxy: open clusters and 
associations: individual: NGC 6940
\end{keywords}

\section{Introduction}\label{intro}

Open star clusters give us important clues to help solve 
remaining puzzles of stellar evolution.
Classical star cluster formation theories assume that members 
of individual open clusters (OCs) were formed from the same chemically 
homogeneous molecular cloud at the same time.
This constraint yields numerous advantages for studies of both stellar
and Galactic disc evolution.
Since we can estimate cluster member stars' ages, masses,  metallicities and 
reddening parameters more reliably than for field stars, 
we can analyze
their model atmospheres and surface abundances more 
accurately.
These data, in turn, lead to better determination of cluster member 
evolutionary status.
In particular, precise positions of red giants (RG)\footnote{
In this paper we will use the term red giant (RG) generically, meaning
all cool and luminous stars; RGB will designate stars on the first ascent 
of the giant branch; and RC will designate He-burning red clump stars.} 
in the color-magnitude diagram (CMD) along with  
the accurate abundances of their light elements
involved in hydrogen burning (Li, C, N, O, $^{12}$C/$^{13}$C) can 
help us to differentiate the life stories of evolved cluster stars.

In standard stellar evolutionary theories, during the evolution from
the main sequence to the tip of the RG branch the only mechanism that
changes the surface abundances is so-called first-dredge up.
The convective envelope movement of red giant branch (RGB) stars 
carries CN-cycle nuclear processed material to the surface. 
This results in depletion in the abundances of Li and C, increase
in N abundance, and a drop in the \ciso\ ratio from 
initial values of about 90 on the main sequence
to about 20-30 (e.g. \citealt{iben67,dear76}).

The outline of this chemical evolution from the main 
sequence to the RG domain is qualitatively well understood; chemical
composition studies of field and OC stars confirm the predicted 
surface abundance changes.
However, there are significant remaining puzzles.
For example, consider the  \ciso\ ratio.
This is one of the most useful tracers of stellar evolution stages; it is 
predicted well by theory and easy to determine observationally.
However it has been long known that many RGs in the field 
and in open or globular clusters have \ciso~$<$~20 (\eg, \citealt{day,tom,
lam81,gil,gilb,luck,tau00, tau01,tau05,smi,mik10,mika, mik11,mik12}).
The isotopic ratios of some RGs even approach the CN-cycle 
equilibrium value of \ciso~=~3$-$4 \citep{cau65}.
These low values are clear signs of further mixing and/or significant
mass loss that is not accounted for by
the classical stellar evolution theories.
The elemental abundances of other CN-cycle participants often are also
at odds with predictions.
The source/theory/mechanism(s) of the extra-mixing is 
an ongoing discussion.

To better understand the discrepancy between classical 
theories and observations in evolving stars, there is increasing attention 
to high-resolution spectroscopic studies of RGs in OCs.
Several groups are increasing the knowledge in this field;
e.g, \cite{dra16,reddy16,smi16}, and references therein.
For nearly two decades the WIYN Open Cluster 
Study\footnote{http://www.astro.ufl.edu/~ata/wocs/} 
\citep[eg.][]{mat}
group has systematically investigated the photometric and spectroscopic
properties of OCs.
Renewed efforts on cluster membership, better spectroscopic data, 
emphasis on gathering ``complete''  RG samples in individual clusters,
and new laboratory transition studies combine to make OCs
attractive targets for chemical composition analyses. 
OCs have typically less than a few thousand members and ages less than several
Gyr, and so each cluster has relatively few RGs to study.

\begin{table}
 \begin{minipage}{75mm}
 \begin{center}
  \caption{NGC 6940 and Hyades cluster parameters.}
  \label{tab1}
   \centering
  \begin{tabular}{@{}lll@{}}
  \hline
   Quantity    &   NGC~6940 & Hyades     \\ 
 \hline
Right Ascension (2000)$^a$   &     20 34 26        &  04 26 54 \\
Declination (2000)$^a$       &   $+$28 17 00       &  $+$15 52 00  \\
Galactic longitude       &      69.860         &  180.064   \\
Galactic latitude        &    $-$7.147         &   $-$22.343 \\
Distance ($\rm {pc}$)         &    770$^c$   &   45$^d$   \\
$E(B-V)^b$                  &     0.214            &   0.01\\
$(m-M)$                  &     10.08$^c$           & 3.30$^a$ \\
log Age                  &     8.86            & 8.90   \\
\hline
\end{tabular}
\end{center}
$^a$ From the WEBDA database.\\
$^b$ \cite{lot01}\\
$^c$ \cite{kha05}\\
$^d$ \cite{mal}\\
\end{minipage}
\end{table} 

The general aim of our project is to present abundance analyses of all 
known non-variable RG members of about 20 OCs.
Our main focus is to determine the LiCNO abundances
and \ciso\ ratios, and to try to evaluate the
evolutionary status of RG members of the OCs.
We are beginning with detailed studies of a couple of OCs with relatively 
large numbers of RGs.
In the first paper of this series, (\citealt{paperI}, hereafter Paper~I)
we analyzed high-resolution spectra of 10 RG members of NGC~752. 
We confirmed memberships of RGs by measuring radial velocities (RV), 
determined model atmospheric parameters, and derived detailed chemical 
compositions.
We also were able to estimate the evolutionary status of individual
RGs by using their photometric and spectroscopic CMD properties along 
with the light element abundances.

\begin{figure}
  \leavevmode
      \epsfxsize=7.2cm
      \epsfbox{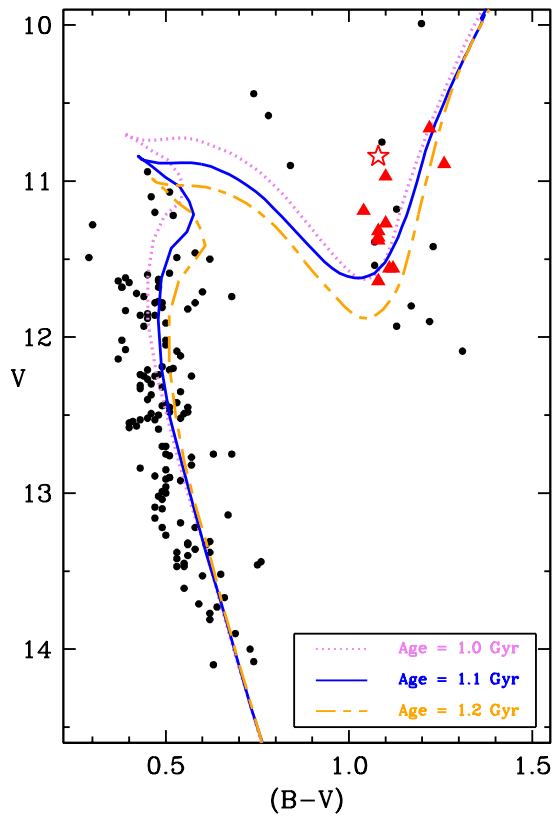}
       \caption{The observed CMD  of NGC~6940 and three
                Yonsei-Yale isochrones. 
                Red triangles are RGs analyzed in this study and the single
                red star is MMU~152 (to be discussed later).
               We also show isochrones for three
                different ages.
                These isochrones have been shifted to match the observed
                CMD by applying the reddening and distance modulus of
                NGC~6940 given in Table~\ref{tab1}. }
     \label{fig1}
\end{figure}

\begin{figure}
  \leavevmode
      \epsfxsize=7.2cm
      \epsfbox{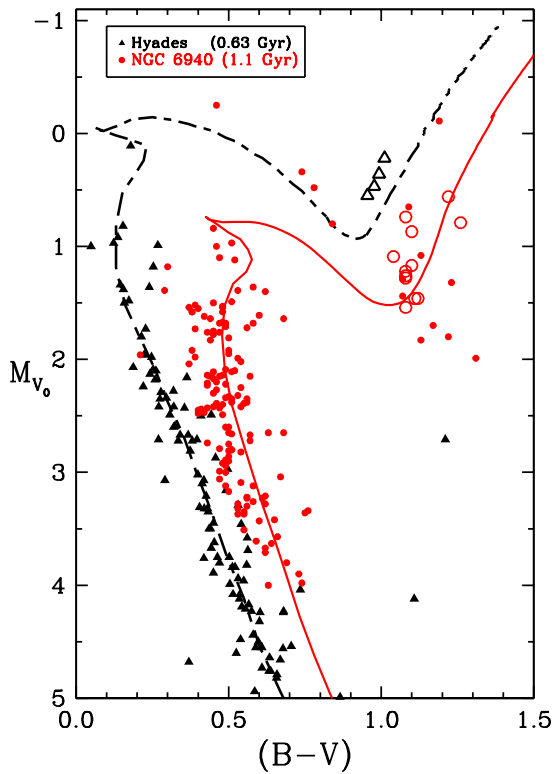}
       \caption{Color-absolute magnitude diagram of the open clusters 
                NGC~6940 and Hyades with Yonsei-Yale isochrones. 
                Open symbols represent the RGs analyzed in this study.}
     \label{fig2}
\end{figure}

In this paper we report on a similar investigation of NGC~6940.
Like NGC~752 of Paper~I, NGC~6940 contains a relatively large population of
RGs: the WEBDA\footnote{http://www.univie.ac.at/webda/webda.html} OC 
database suggests that there are $\sim$20 RG members of this cluster.
NGC~6940 has been mostly neglected in the literature. 
To our knowledge this work presents the chemical composition analysis 
of the first large sample that includes 12 purported RG members of NGC~6940.
We have gathered high resolution, high signal-to-noise spectra. 
We report RVs, atmospheric parameters, [Fe/H] metallicities\footnote{
For elements A and B, [A/B] = log $(N_{A}/N_{B})_{\star}$ 
-- log $(N_{A}/N_{B})_{\sun}$
and log $\epsilon$(A) = log $(N_{A}/N_{H})$ + 12.0 .
Also, metallicity will be taken to be the [Fe/H] value.}, 
and detailed abundances of the $\alpha$,  light odd-Z, Fe-peak, and 
neutron-capture groups, with a particular focus on the LiCNO elements.
We also include abundance analysis of the four well-known
RG members of the Hyades open cluster in order to standardize our
analysis of NGC~6940.

The structure of this paper is as follows: 
in \S\ref{cluster6940} we summarize the history of 
NGC~6940 and its RGB members.
The observations and reductions are outlined in \S\ref{obs}.
We discuss compilation of atomic/molecular line lists and equivalent width 
measurements in \S\ref{lines}.
The derivation of model atmospheric parameters is described in \S\ref{models},
followed by abundance analysis in \S\ref{abunds}.
Finally, in \S\ref{discuss} we discuss the implications of our results
for the evolution of NGC~6940.

\section{NGC~6940}\label{cluster6940}

NGC~6940 is an intermediate age ($\approx$1 Gyr) open 
cluster at a distance of $d$~$\approx$~800 pc; overall cluster parameters 
are gathered in Table~\ref{tab1}.
NGC~6940 has probably not been studied extensively because 
its members are not easily distinguished from field stars along the 
same sight line.
\cite{tru} classified this cluster as  \rm{III 1m}, meaning that it is
a detached cluster with no noticeable central concentration, containing
50-100 stars of nearly the same apparent magnitude.
The star numbering system of NGC~6940 is mostly based on \cite{vas}, as
reorganized by \cite{mer08}. 
We have adopted this numbering system, adding the prefix ``MMU'' for
star names in this paper.

\subsection{Previous Studies}\label{previous}

Two early studies of OC RGs included NGC~6940.
Using \cite{wal} data, \cite{hart} studied NGC~6940 as part of an investigation 
focused on
$E(U-B)/E(B-V)$ color excess ratio, and found 
fairly uniform reddening.
Assuming a commonly used color excess ratio of 0.72, they derived 
$E(B-V)_{ms}=0.20\pm0.005$, $E(B-V)_{giants}=0.18\pm0.012$.
\cite{jen} derived metallicities, color excesses and distance moduli 
for 20 clusters by observing their K-giants with UBViyz photometry. 
They observed 20 RGs in NGC~6940 and calculated the following 
quantities: [Fe/H] = $-0.2$, $E(B-V) = 0.16$, 
$(m-M)_{0} = 10.2$ and age $t$~=~$5\times10^{8}~yr$.  

The first relative proper motion study of NGC~6940 to 
estimate cluster membership was conducted by \cite{vas}.  
The cluster drew some attention in this era because it contains a 
semi-regular M0 giant (FG Vul) and it appears to have a peculiar 
RGB with variable reddening. 
Because of these reasons it has been subject to many photographic and 
photometric studies in UBV system such as \cite{wal}, \cite{lar60}, 
\cite{hoag}, and \cite{john}.
Various photometric and membership studies were cross-correlated by 
\cite{lar64}, who concluded that interstellar reddening is irregular 
across the cluster face 
($E(B-V)$~=~0.05 to 0.30, \cite{tag}.)

\cite{geis} has conducted one of the first extensive 
radial velocity ($RV$) study of NGC~6940. \citeauthor{geis} derived the mean
cluster RV, $\langle$RV$\rangle$ = $5.7 \pm 1.5\ {\rm km\ s^{-1}}$, 
by gathering medium-resolution spectroscopy of 11 stars.     
\cite{mer89} conducted the RV study of NGC~6940, 
using observations of 24 possible member RGs.
Rejecting four of the targets as non-members, they derived a mean cluster 
velocity of $\langle$RV$\rangle$ = $7.75 \pm 0.13\ {\rm km\ s^{-1}}$ from the 
20 RG members, and found that six of them are spectroscopic binaries. 
Recently, \cite{mer08} observed 26 possible NGC~6940 RGs, finding five 
of them to be non-$RV$ members.
They also updated their previous RV measurements and derived a cluster
velocity of $\langle$RV$\rangle$ = $7.89 \pm 0.14\ {\rm km\ s^{-1}}$. 
Additionally, they discussed the morphology of the cluster RGB in general, 
noting in particular star MMU~152, one of our chosen targets.
This RG appears to have the CMD location of a typical binary RG, but it did not show any variability in its velocity data.
Therefore, they concluded that MMU~152 is possibly a double star with 
large separation.

  \begin{table*}
 \centering
 \begin{minipage}{180mm}
 \begin{center}
  \caption{Basic parameters of observed NGC~6940 RGs.} 
  \label{tab2}
  \begin{tabular}{@{}lccccccccccc@{}}
  \hline
   Star &    RA & DEC  &  $B^a$ &  $V^a$  &  $K^b$  & $M_{V_0}$ & $(B-V)$  & $(B-V)_{0}$& $(V-K)_{0}$ &$\mu_\alpha$  &  $\mu_\delta$       \\ 
    & (2000)   &  (2000)   & && && && &   (mas/yr)  &  (mas/yr)   \\               
 \hline
MMU~28$^c$	    &	 20 33 25.016 &	$+$28 00 46.97	&	12.68	&	11.56	&	8.93&1.46	&1.12& 0.91 & 2.42 & $-$5.70 &$-$14.6\\
MMU~30$^c$ 	    &	 20 33 29.806 	&	$+$28 17 05.07	&	12.15   &	10.89	&	8.18&0.79	&1.26& 1.05  &2.50 & $-$5.70&$-$7.6     \\
MMU~60$^d$   	& 	 20 33 59.570	&	$+$28 03 01.64	&	12.67	&	11.56	&	8.97	&1.46	&1.11& 0.90 & 2.38&$-$3.4 & $-$8.1      \\
MMU~69$^d$ 	    & 	 20 34 05.746 	&	$+$28 11 18.38	&	12.72   &	11.64	&	8.95	&1.54	&1.08& 0.87 &2.48&$+$1.5 & $-$9.6      \\
MMU~87$^d$ 	    &	 20 34 14.693 	&	$+$28 22 15.76	&	12.40	&	11.32	&	8.74 &1.22	&1.08& 0.87&2.37&$-$3.3  &  $-$11.0   	\\
MMU~101$^d$    &	 20 34 23.645 	&	$+$28 24 25.62	&	12.37	&	11.27	&	8.74	&1.17	&1.10& 0.89 &2.32&$-$2.6  &  $-$8.7 	 	\\
MMU~105$^c$ 	&	 20 34 25.458 	&	$+$28 05 05.62  &	11.88	&	10.66   &	7.82 &0.56   &1.22& 1.01 & 2.63 &$-$5.0& $-$13.0    \\
MMU~108$^c$ 	&	 20 34 25.675 	&	$+$28 13 41.54  &	12.23	&	11.19	&	8.70	& 1.09  &1.04& 0.83&2.28&$-$3.4   & $-$11.4 	   	 \\
MMU~132$^c$ 	&	 20 34 40.112 	&	$+$28 26 38.94	&	12.07	&	10.97   &	8.40 &0.87	 &1.10&  0.89 & 2.37 &$-$6.8 & $-$17.6 \\
MMU~138$^d$ 	&	 20 34 45.875 	&	$+$28 09 04.62	&	12.44	&	11.36   &	8.81	&1.26	 &1.08& 0.87&2.34&$-$4.8    & $-$12.8	 \\
MMU~139$^d$ 	&	 20 34 47.610 	&	$+$28 14 47.25	&	12.46 	&	11.38   &	8.80	&1.28	 &1.08& 0.87&2.37&$+$0.1   &   $-$7.5     \\
MMU~152$^{c,d}$ 	&	 20 34 56.637 	&	$+$28 14 27.27	&	11.92 	&	10.84	&	8.27 &0.74	 &1.08&  0.87 &2.36 &$-$1.9 &$-$4.6 	       \\
 \hline
\end{tabular}
\end{center}
$^a$ \cite{lar60} \\
$^b$  \cite{2MASS}\\
$^c$  Coordinates and proper motions from \cite{hog}\\
$^d$ Coordinates and proper motions from\cite{hog00}\\
\end{minipage}
\end{table*}

\cite{stro1} calculated  ``absolute'' metallicities of 72 OCs from 
previous spectroscopic and photometric studies,
obtaining [M/H] = $-0.1$ for 
NGC~6940 with probable error not larger than 0.1 dex. 
In a companion paper, \cite{stro2} suggested that [M/H] = $-0.14$ 
from consideration of the age and Galactic position of NGC~6940.
The mean of more recent metallicitiy estimates generally agree on
a solar metallicity for this cluser, albeit with some scatter.
\cite{tho} obtained moderate-resolution CCD spectra and measured [Fe/H] 
values via Mg and Fe spectroscopic indices, finding 
[Fe/H] = $-0.06 \pm 0.13$ for NGC~6940 from six RG members.
The \cite{dia02} OC
compendium\footnote{https://heasarc.gsfc.nasa.gov/W3Browse/all/openclust.html}
recommends [Fe/H]=~$+0.04\pm0.10$.
\cite{twa} included NGC~6940 in their investigation of photometric and
spectroscopic data for 76 open clusters, deriving [Fe/H]=~$0.01\pm0.06$. 
\cite{friel} calculated metallicity by using the spectroscopic data 
from \cite{tho}, finding [Fe/H]=~$-0.12\pm0.10$.
By using \cite{hoag} photometric data \cite{cam}
calculated [Fe/H]=0.014 from transformed ultraviolet excesses. 
Recently, \cite{bla} included one NGC~6940 RG in their high-resolution
``chemical tagging'' spectroscopic study of 31 OCs.
Abundances of 17 species and [Fe/H] was 
derived for this one RG member of NGC~6940 from the NARVAL 
spectropolarimeter  (R $\cong$ 81,000) data. 
They calculated metallicities from two slightly different 
normalization processes: [Fe/H]$_{1}$ = $+0.04$ and 
[Fe/H$]_{2}$ = $+0.09\pm0.07$.
From all of these studies we conclude that NGC~6940 has
near-solar overall metallicity.

\subsection{Colour-Magnitude diagram and RG target selection}\label{selection}

Photometric and photographic observations of NGC~6940 have 
been done by several authors (e.g. \citealt{wal,lar60,hoag,john,lar64,ste}). 
To our knowledge, there is no recent homogeneous photometric 
survey that could create a comprehensive CMD that includes our target RGs.
Therefore we chose to use 
photographic colors and magnitudes
compiled from the sources provided in WEBDA database. 
In Figure~\ref{fig1} we have plotted the data from \cite{lar60}.
These data contain 164 probable members of NGC 6940. 
Sanders (1972) reported astrometric cluster membership probabilities of 216 stars in the field of the NGC 6940. 
115 stars in \cite{lar60} list are confirmed as members of this cluster by \cite{san}.
The rest of the \citeauthor{lar60} sample are not included in  \citeauthor{san}'s study. 
All our RGs are confirmed 
as the members of the NGC 6940  open cluster in \cite{san,mer89} and \cite{mer08}.
Also 10 of the RGs we studied here have been presented
 in the catalog of membership probabilities published by  \cite{dia14}  and they all have membership probabilities over 90\%. 
In Figure~\ref{fig1} we also show three isochrones from the 
latest set of the Yonsei-Yale (Y$^{2}$) project \citep{YY}\footnote{
http://www.astro.yale.edu/demarque/yyiso.html}. 
To match the observations we applied the reddening and distance modulus
values given in Table~\ref{tab1}, and have assumed $A_v$ = 3.1$E(B-V)$. 
Our main goal was to estimate the turn-off mass for the cluster and use this
information to have a better estimate on the model atmospheric 
parameters of the cluster members (see \S\ref{initial}). 
The best average match between the NGC~6940 observed CMD 
and Y$^{2}$ isochrones is for an age log$(t)=9.04$ (1.1 Gyr), 
somewhat older that the WEBDA value of 8.86, and for a
metallicity Z~=~0.016 (equivalent to [M/H]~=~$-0.06$).
We also tried matching other sets of isochrones to the data
besides Y$^{2}$, such as PARSEC \citep{bre12} and Dartmouth \citep{dot08},
finding that Y$^{2}$ isochrones provide the best-fitting model 
isochrone to the cluster data.
However, the different isochrone grids all yielded 
similar results, suggesting
 a turn-off mass range from 1.9 to 2.1 M$_{\odot}$ for NGC~6940.
Therefore we have decided to adopt an average turn-off mass of 
$M_\textrm{TO}$=2~M$_{\odot}$ for this cluster.

The WEBDA age of NGC~6940 listed in Table~\ref{tab1},
log$(t)=8.86$ was adopted from \cite{lot01}. 
\cite{lyn95}\footnote{http://cdsarc.u-strasbg.fr/cgi-bin/myqcat3?VII/92A} 
recommended log$(t)=9.04$, taking their value from \cite{van80}. 
\cite{kha05} reports log$(t)$ = 8.94 for the cluster. 
We also applied Y$^2$ isochrones to the well-known open cluster Hyades 
to compare the two clusters, thus gaining more confidence in the age 
estimate of NGC~6940. 
The Hyades age is very well determined, log$(t)$ = 8.80 (625 $\pm$ 25 Myr, 
\citealt{perry98}), as is its main sequence turn-off mass, 
2.3 M$_{\odot}$ \citep{weid}.
Using the photometric data of Hyades taken from \cite{john55}, we compare the 
isochrone fittings of both clusters in Figure~\ref{fig2}. 
As in Figure~\ref{fig1} we added extinction to the
(B-V) isochrones in Figure~\ref{fig2}, to separate clearly the NGC~6940 
and Hyades OC sequences. 
We have transformed $V$ into $M_{V_0}$ using the parameters of 
Table~\ref{tab1}, again assuming $A_v$ = 3.1$E(B-V)$ .
It is clear that NGC~6940 is much older than Hyades, and we accept
the age implied by our Y$^{2}$ isochrone fit, log$(t)$ = 9.04.

Further refinement of cluster membership was accomplished in 
the RV survey of \cite{mer08}, who observed 26 RG possible
member of NGC~6940 with CORAVEL ``spectrovelocimeter''.  
They derived $\langle$RV$\rangle$ = $7.89\pm0.14$ for NGC~6940, and 
used that mean velocity
to eliminate five probable non-members, paring the list to 21 RGs.
Our target list contains 13 stars from that 
study, after the elimination of
an additional eight stars suggested to be spectroscopic binaries or
have variable velocities in the \citeauthor{mer08} survey.
We were able to gather high resolution spectra of 12 of 
these member RGs.  
Program stars are listed in Table~\ref{tab2} by their MMU 
designations, along with their coordinates and proper motions from
SIMBAD\footnote{http://simbad.u-strasbg.fr/simbad/},
$B$ and $V$ magnitudes from \cite{lar60}, and $K$ magnitudes 
from 2MASS \citep{2MASS}. 
Table~\ref{tab2} also contains 
observed $(B-V)$ color indexes, de-reddened $(B-V_0)$
and $(V-K_0)$ values, and and absolute magnitudes $M_{V_0}$.
Our NGC~6940 RGs are marked with red triangles and star 
in Figure~\ref{fig1}.  
These occupy a very small CMD domain:  0.85~$\leq$~$B-V$~$\leq$~1.05,
0.5~$\leq$~$M_V$~$\geq$~1.

\section{\textbf{Observations and Reductions}}\label{obs}

High-resolution ($R\equiv\lambda/\Delta\lambda\approx$ 60,000) 
spectra of target RGs were gathered with the 
McDonald Observatory 9.2m Hobby-Eberly Telescope (HET) High-Resolution 
Spectrograph (HRS; \citealt{tull}).
This instrument achieves a large spectral coverage from a mosaic of
two CCD detectors.
In the configuration adopted for this project the spectral coverage is 
5100$-$6900~\AA\ on the ``blue'' detector and 
7000$-$8800~\AA\ on the ``red'' detector.
Observations were conducted between April and June of 2013.
Individual exposure times and S/N ratios are given in Table~\ref{tab3}.  
Two exposures were taken for each target.
Our stars are solar-metallicity RGs and thus have very 
line-rich spectra.
Therefore it is very difficult to find pure continuum regions to specify 
$S/N$ ratios from the reduced spectra.
The $S/N$ ratios in Table~\ref{tab3} are taken from the observing night 
reports of the HET. 

After splitting the raw data images into red and blue sections, 
we applied standard reduction procedures for both sections by 
using
IRAF\footnote{http://iraf.noao.edu/} routines.
We started with overscan removal, bias subtraction, 
flat-fielding and scattered light subtraction with the tasks in the 
\textit{ccdred} package. 
Then commands of the \textit{echelle} package were used to extract the
single-order spectra.
We used the \textit{ecidentify} task for wavelength calibration, based on
ThAr lamp exposures taken each observing night.
Removal of the telluric (atmospheric) lines was conducted with 
\textit{telluric} task, which divides out the telluric lines from the 
spectrum of the program star by using the spectrum of a hot, rapidly 
rotating (telluric) star obtained in the same observing night.

High resolution (R $\approx$ 60.000) observations 
of four RGs of the Hyades open cluster were conducted with the 
Robert G. Tull Cross-Dispersed Echelle Spectrograph \citep{tull95} 
on the 2.7 m Harlan J. Smith Telescope at McDonald Observatory in October 2012.
The spectral coverage of this setup is 4000$-$8000 \AA\ and we 
applied the same reduction procedure as described above.

To measure the RVs of the OC members we used synthetic spectra
that we created using the same line list   
(described in \S\ref{lines}) as templates. 
These templates have similar \teff, \logg, and [Fe/H] to our program stars. 
We determined four clear spectral orders that have as many stellar lines 
as possible without much telluric contamination both in blue and red regions, 
and then created template synthetic spectra for the 
wavelength regions of 8290$-$8430 \AA, 7960$-$8090 \AA, 7390$-$7505 \AA, 
7030$-$7160 \AA\ in the red, and 6720$-$6840 \AA, 5990$-$6090 \AA, 
5710$-$5800 \AA\ 5260$-$5350 \AA\ in the blue. 
Heliocentric corrections were made with the \textit{rvcorrect} task in IRAF. 
Then the RV values were calculated for each region
using the cross-correlation task \textit{fxcor} \citep{ftz93}.
In Table~\ref{tab3}, we list the final RV values along with their standard 
deviations ($\sigma$) for each star, which we achieved by taking the average 
of RVs measured from individual regions.

The difference between the average of \textit{fxcor} errors and standard 
deviations of eight RVs vary from star to star in the range 
0.05$-$0.10~\kmsec.
The mean RV of the cluster, $\langle$RV$\rangle$~=~8.02~$\pm$~0.16~\kmsec\ 
($\sigma=0.56$), agrees 
with the \cite{mer08} value ($\langle$RV$\rangle$~=~7.89~$\pm$~0.14~\kmsec) 
within mutual uncertainties.

Table~\ref{tab3} also contains RVs of each star reported by \cite{mer08}. 
According to our RV determinations MMU~28 deviates 
from the cluster mean by $+$0.88~\kmsec\ ($\sigma$~=1.5~\kmsec) 
and MMU~152 by $+$1.26~\kmsec. 
These RVs may lower their stars' cluster membership probabilities slightly but 
membership in NGC~6940 is not ruled out. 
The proper motion membership probabilities given for MMU~28 and
MMU~152 by \cite{san} are 60 and 94\%, respectively. 
\cite{dia02} report only the proper motion membership probability 
for MMU~152 and it is
also over 90\%. RV membership probabilities were discussed by \cite{mer89} and \cite{mer08}, they both confirmed the membership status of both stars.

If we combine our values with 
those of \cite{mer08} we conclude that all the RGs we 
have observed are members of NGC~6940 with high probabilities.

\begin{table}
 \begin{center}
 \begin{minipage}{80mm}
  \caption{Exposure times and radial velocities of the observed stars.
  \label{tab3}} 
  \begin{tabular}{@{}lccccc@{}}
  \hline
Star                                &   
Exp.                                &  
S/N                                 &  
$RV^a$                              & 
$\sigma$                             &
$RV^b$                              \\ 
    &   (\textit{s})  &  @6948 \AA   &(km s$^{-1}$)  &&(km s$^{-1}$)   \\               
 \hline
MMU~28 	    &	  	1000    &	130	&	8.90$\pm$0.08	& 0.22 &	 7.99$\pm$0.16  \\
MMU~30 	    &	  	700  	&	113     &	7.96$\pm$0.07	& 0.20&	 7.63$\pm$0.15  \\
MMU~60      & 	  	1050 	&	121	&	7.66$\pm$0.08	& 0.22&	 7.27$\pm$0.18  \\
MMU~69 	    & 	 	1050	&	93      &	8.08$\pm$0.08	&0.24&	  7.56$\pm$0.15 \\
MMU~87 	    &	 	900	&	103	&	7.98$\pm$0.09	&	0.27&  7.45$\pm$0.16 \\
MMU~101     &	 	900	&	155	&	7.74$\pm$0.08	&	0.23&  6.81$\pm$0.14 \\
MMU~105     &	 	650     &	116	&	7.74$\pm$0.08   &0.23& 7.58$\pm$0.13  \\
MMU~108     &	 	900     &	145	&	7.39$\pm$0.09	&0.25&	 6.76$\pm$0.13  \\
MMU~132     &	 	700	&	129	&	7.76$\pm$0.15   &0.42&	 7.17$\pm$0.14  \\
MMU~138     &	 	900	&	82	&	8.22$\pm$0.08   &0.23&	 7.55$\pm$0.15  \\
MMU~139     &	 	900   	&	136 	&	7.53$\pm$0.08  &0.23&	 7.12$\pm$0.16  \\
MMU~152     &	 	650	&	144 	&	9.28$\pm$0.08	&0.24&	 8.50$\pm$0.15  \\
 \hline
\end{tabular}
\end{minipage}
\end{center}
$^a$ this study   \\
$^b$ \cite{mer08} \\
\end{table}

\section{\textbf{LINE LISTS AND EQUIVALENT WIDTHS}}\label{lines}

To evaluate the atmospheric parameters and chemical abundances of all species 
we used the current version of the LTE line 
analysis and synthetic spectrum code MOOG\footnote{
http://www.as.utexas.edu/~chris/moog.html} 
\citep{sne73}.
Two techniques were employed to derive the chemical abundances: 
equivalent width ($EW$) measurement and spectral synthesis. 
Measured $EW$s of unblended neutral and ionized Fe and Ti were used to 
calculate the atmospheric parameters. 
Abundances from \mbox{Si\,{\sc i}}, \mbox{Ca\,{\sc i}}, 
\mbox{Cr\,{\sc i}}, \mbox{Cr\,{\sc ii}} and \mbox{Ni\,{\sc i}} transitions
were also derived from $EW$ measurements.
The other atomic and molecular species, which have 
spectral lines that are blended and/or afflicted by hyperfine/isotopic 
substructure, were analyzed using the spectral synthesis technique.

\subsection{Line Lists\label{linelists}}

\begin{table}
 \centering
 \begin{minipage}{75mm}
  \caption{Atomic transitions used in this study.
  The machine-readable version of
  the entire table is available in the online journal.}
  \label{tab4}
  \begin{tabular}{@{}lccccc@{}}
  \hline
Species   &  Wave.  &  LEP  &  log \textit{gf}  &  EW / syn   & Ref.   \\
      &   (\AA)  & (eV)   &  &  &   \\
\hline
\mbox{Li\,{\sc i}}	    &	6707.9    &	0	&           $0.17$	&	syn	& Kurucz \\
CH	                       &	4310	   &	 	&	                    	&	syn	 & Mas14\footnote{\cite{mas14}} \\
CH                         	&	4325       &	 	&	                       &	syn	 & Mas14 \\
C$_{2}$	                &	5160       &	 	&  	                   &	syn	 & Bro14\footnote{\cite{brooke14a}}\\
C$_{2}$	               &	5630       &	 	&	                       	&	syn	 & Bro14\\
CN	                       &	8000       &	    &	                     	&	syn	 & Sne14\footnote{\cite{sne14}}\\
\mbox{O\,{\sc i}}      &	6300.31	&	0	        &	$-9.72$ 	&	syn	 & Allen01\footnote{\cite{allen01}}\\
\mbox{Na\,{\sc i}}	&	5682.64	&	2.101	&	$-0.71$	&	syn	 & NIST\\
\mbox{Na\,{\sc i}}	&	6154.23	&	2.101	&	$-1.55$	&	syn	 & NIST\\
\mbox{Na\,{\sc i}}	&	6160.75	&	2.103	&	$-1.25$	&	syn	 & NIST\\
\mbox{Mg\,{\sc i}}	&	5528.41	&	4.343	&	$-0.62$	&	syn	  & Kurucz\\
\mbox{Mg\,{\sc i}}	&	5711.08	&	4.343	&	$-1.83$	&	syn	 & Kurucz\\
\mbox{Mg\,{\sc i}}	&	7811.11	&	5.941	&	$-0.95$	&	syn	 & Kurucz \\
\mbox{Al\,{\sc i}}	    &	6696.02	&	3.14	&	  $-1.35$	   &	syn	 & Kurucz \\
\mbox{Al\,{\sc i}}	 	&	6696.18	&	4.018	&	$-1.58$ 	&	syn	& Kurucz \\
\mbox{Al\,{\sc i}}	 	&	6698.67	&	3.14	&	 $-1.64$  	&	syn	 & Kurucz\\
\mbox{Al\,{\sc i}}	 	&	7835.30	&	4.018	&	 $-0.65$	&	syn	  & Kurucz\\
\mbox{Si\,{\sc i}}	 	&	5488.98	&	5.614	&	$ -1.90$	&	EW	 &Lob11\footnote{\cite{lob11}}\\
\mbox{Si\,{\sc i}}		&	5517.53	&	5.082	&	$-2.61$ 	&	EW	  &VALD\\
\mbox{Si\,{\sc i}}		&	5665.55	&	4.92	&	 $-2.04$	    &	EW	 &NIST\\
\hline
\end{tabular}
\end{minipage}
\end{table}

In Paper~1 on OC NGC~752 we generated a line list of 
relatively unblended lines that have reliable transition probabilities 
in the yellow-red spectral region.
For the present study we used a slightly updated line list with some 
addition of \mbox{Fe\,{\sc i}} and \mbox{Ni\,{\sc i}} lines.
The complete line list, which includes both these EW 
transitions and those later to be used in spectrum syntheses, is presented 
in the online version of the Table~\ref{tab4}, which contains wavelengths 
$\lambda$, lower excitation energies $\chi$, transition probabilities 
log $gf$, and references for the transition probabilities.

\begin{table*}
 \centering
 \begin{minipage}{140mm}
  \caption{Equivalent width measurements (m\AA) of the first five lines for
  the RGs of the NGC~6940. The machine-readable
  version of the entire table is available in the online journal.
  \label{tab5}}
  \begin{tabular}{@{}lccccccccccccc@{}}
  \hline
  Species  &  Wavelength  &   \multicolumn{12}{c}{MMU}  \\
  &(\AA)& 28 &  30  & 60  & 69  & 87  & 101  & 105  & 108  & 132  & 138 & 139 & 152  \\    
\hline
\mbox{Si\,{\sc i}}  & 5488.98 &31.3&       &32.1&31.6&      &31.7&    &28.2&40.0&       &35.1&\\
\mbox{Si\,{\sc i}}  & 5517.53 &22.5&29.6&25.5&27.6&       &      &    &21.2&30.1&25.7&24.3&29.4\\
\mbox{Si\,{\sc i}}  & 5665.55 &64.4&71.9&       &63.4&       &66.0&    &61.7&73.4&62.5&67.6&      \\
\mbox{Si\,{\sc i}}  & 5690.43 &61.9&69.1&61.0&62.7&72.5& 65.4&   &62.3&       &63.5&       &   \\ 
\mbox{Si\,{\sc i}}  & 5772.15 &68.3&       &       &       &        &75.2&    &65.6&80.2&      &71.3&77.3\\
\hline
\end{tabular}
\end{minipage}
\end{table*}

The chosen wavelength range of $EW$ measurements was 5200$-$7100 \AA.
We tried to avoid regions of heavy telluric line contamination, and did 
not explore the crowded bluer spectral regions ($\lambda$~$<$~5100~\AA)
since the determination of their continua is difficult.

We derived the model atmospheric parameters from 
71 \mbox{Fe\,{\sc i}},  12 \mbox{Fe\,{\sc ii}}, 11 \mbox{Ti\,{\sc i}} and 
4 \mbox{Ti\,{\sc ii}} lines; the numbers vary from star to star.
After several trials we put some restrictions to the Fe and 
Ti transitions that were used to calculate the model atmosphere parameters.
In particular, we discarded
very weak and strong lines for these species
by limiting the $EW$s to 10~m\AA~$<$~EW~$<$~150~m\AA,
or $RW$~$=$ $log~(EW/\lambda)$~$\sim$~$-$5.8 to $-$4.6 at
$\lambda$~$\simeq$~6400~\AA. 
We did not apply the $EW$ limitation to other species.        

We have adopted various references for the transition probabilities. 
When possible we tried to use laboratory-based homogeneous 
(single source) $gf$ and isotopic/hyperfine structure data. 
Unfortunately, there are no recent 
comprehensive lab studies for the important species 
\mbox{Fe\,{\sc i}} and \mbox{Fe\,{\sc ii}}, so for them
we had to use several sources (e.g. \citealt{obr91}, \citealt{den14},
\citealt{ruf14}, NIST\footnote{
http://physics.nist.gov/PhysRefData/ASD/lines\_form.html},
and VALD\footnote{
http://vald.inasan.ru/~vald3/php/vald.php}
\citep[eg.][]{rya}).
The useful spectral range of the HET data for our OC work 
is a little different from that of the McDonald 2.7m data, so we added 
twenty more \mbox{Fe\,{\sc i}} lines to the list used in Paper~1.
Most of the new lines are from \citeauthor{den14} and \citeauthor{ruf14}. 
For the species \mbox{Ti\,{\sc i}}, \mbox{Ti\,{\sc ii}}, 
\mbox{V\,{\sc i}},
\mbox{Cr\,{\sc i}}, \mbox{Co\,{\sc i}}, \mbox{Ni\,{\sc i}}, 
\mbox{La\,{\sc ii}}, \mbox{Nd\,{\sc ii}} and \mbox{Eu\,{\sc ii}} 
we exclusively used lab data obtained by the University of Wisconsin 
atomic physics group.

For the present study, we investigated some of the 
potentially blended lines more closely and updated the 
$gf$s of a few lines as well as the number of lines used for the abundance 
determinations for some species. 
All of these changes were first tested on the very high-resolution solar 
\citep{kur84} and Arcturus flux spectra \citep{hink00}. 
Line lists for a few individual species deserve comment here.

\textit{Manganese:}
As in Paper~1, we determined manganese abundances from the 
\mbox{Mn\,{\sc i}} the triplet around 6016 \AA\ region. 
We used synthetic spectra due to the many hyperfine structure ($hfs$) 
components that make up each of these transitions.
Investigation of the 6016.7 \AA\ \mbox{Mn\,{\sc i}} line profile using both 
solar and Arcturus spectra showed that a \mbox{Fe\,{\sc i}} line at 
6016.61 \AA\ is blended with the Mn $hfs$ and creates an unrealistic 
asymmetric line profile around the \mbox{Mn\,{\sc i}} components,
which in turn affects the Mn abundance determined from this line. 
Proper accounting of the \mbox{Fe\,{\sc i}} blend allowed us to measure 
internally consistent Mn abundances from the triplet around 6016 \AA.

\textit{Copper:}
In this study, we used only the 5782.13 \AA\ \mbox{Cu\,{\sc i}}
transition, adopting the overall $gf$ value and the isotopic and hyperfine 
splitting from \cite{cun02}, who used the laboratory data of \cite{han83}. 
We assumed solar-system isotopic percentages of $^{63}$Cu (69.2\%) and
$^{65}$Cu (30.8\%).
We were not able to use other \mbox{Cu\,{\sc i}} lines located at 5218.2 and 
5220.06 \AA\ due to severe atomic and molecular line contamination. 
The prominent 5105.5 \AA\ line lies outside our HET HRS 
spectrum coverage.

\textit{Yttrium and Neodymium:}
We used only \mbox{Y\,{\sc ii}} lines for this element.
We added four more lines to the list used in Paper~I, but we had to drop 
5087.42 \AA\ from the present study; it is also beyond the HRS coverage.
We also added two more lines in the analysis of \mbox{Nd\,{\sc ii}}.

\textit{Scandium, Vanadium, and Cobalt:}
In Paper~I, we had to give special treatment to the Fe-group 
species of \mbox{V\,{\sc i}}, \mbox{Co\,{\sc i}} and \mbox{Sc\,{\sc ii}} 
transitions since they have significant hyperfine substructure and 
lacked recent lab studies.
Their abundances were computed from $EW$s input to 
MOOG's blended-line option \textit{blends} after deriving empirical
transition probabilities from the solar spectrum.
For the present work, we were able to adopt new 
\mbox{V\,{\sc i}} laboratory data from \cite{law14}, and new lab 
\mbox{Co\,{\sc i}} data from \cite{law15}.
These fresh line data include improved hyperfine substructure parameters.
No lab studies for \mbox{Sc\,{\sc ii}} have been 
published recently, so we used reverse solar to calculate their 
line transition probabilities.
That is, we measured $EW$s from a very high-resolution solar flux spectrum 
\citep{kur84} and forced their $gf$s to give the solar abundances 
recommended by \cite{asp09}. 
We then used these calculated $gf$s in similar calculations for the
program stars.

Since later in the paper we compare the abundances of NGC~6940 and Hyades 
clusters with NGC~752 results, we applied these new updates to the 
fresh analyses of NGC~752 RG stars.
We will discuss these new values (Table~\ref{tab12}) in \S\ref{discuss}.
The remaining species 
(C$_2$, CN, [\mbox{O\,{\sc i}}],
 \mbox{Li\,{\sc i}}, \mbox{Na\,{\sc i}}, \mbox{Mg\,{\sc i}}, 
\mbox{Cu\,{\sc i}}, \mbox{Zn\,{\sc i}}, \mbox{Y\,{\sc ii}},
\mbox{La\,{\sc ii}}, \mbox{Nd\,{\sc ii}}, \mbox{Eu\,{\sc ii}})
have complex transition substructures and/or are blended 
with various atomic and/or molecular lines.
We derived their abundances by spectrum synthesis with the MOOG 
\textit{synth} option.

\subsection{Equivalent width measurements}

We used the same Interactive Data Language (IDL) code 
\citep{roe10, bru11} to measure the $EW$s that 
was described in Paper~I.
This interactive code automatically matches the observed lines with theoretical 
Gaussian line profiles or Voigt profiles for some of the stronger lines.
Line depths were also recorded to estimate the initial effective temperatures 
\teff\ (\S\ref{models}) using 
the line depth ratio (LDR) method,
to be described in \S\ref{initial}.
We present $EW$ measurements of our 12 NGC~6940 RGs
in Table~\ref{tab5}.

\section{Model Atmospheres}\label{models}

We made use of the $EW$s of unblended neutral and ionized Fe and Ti lines 
to determine the model atmospheric parameters 
(\teff, log~$g$, $\xi_{t}$, [M/H]) of our program stars. 
Fe and Ti abundances were processed simultaneously 
with the semi-automated version of MOOG. 
Each atmospheric parameter is connected to the abundances of these species 
as follows:
(\textit{i}) for \teff, abundances of low- and high-excitation potential 
($\chi$) lines of \mbox{Fe\,{\sc i}} and \mbox{Ti\,{\sc i}} need to be 
the same on average;
(\textit{ii}) for $\xi_{t}$, there should be no apparent trend between
the reduced width ($RW = log(EW/\lambda)$) and abundances of  
neutral weak and strong lines of \mbox{Fe\,{\sc i}} and \mbox{Ti\,{\sc i}}; 
(\textit{iii}) for log~$g$, mean abundances of neutral and ionized Fe and Ti 
lines need to be in agreement;
(\textit{iv}) and for [M/H], derived metallicity should agree with the 
one assumed in creating the model atmosphere.

In estimating model parameters we applied the same weights 
to Fe and Ti species that were described in in Paper I. 
That is, we gave 0.65 weight to Fe lines, and 0.35 for Ti lines.
MOOG first calculated the abundances from the individual line EWs with 
an initial model atmosphere. 
Then it iteratively altered the input model parameters in response to
the abundance slopes 
with $\chi$ and $RW$, until a balance was achieved between the neutral 
and ionized species for a final model atmosphere.

\subsection{Initial parameters}\label{initial}

\begin{figure}
  \leavevmode
      \epsfxsize=8cm
      \epsfbox{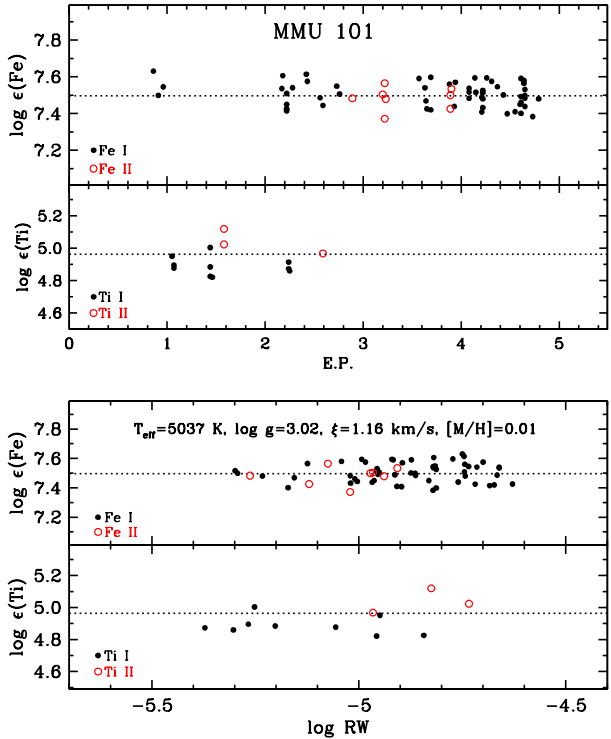}
       \caption{An example model atmosphere parameter
                determination using both ionized and neutral species of 
                Fe and Ti for MMU 101.
                In the top panel the line abundances are plotted as a
                function of excitation potential $\chi$ (E.P.), and in the
                bottom panel as a function of reduced width RW.
                Iterating T$_{{\rm eff,spec}}$, \logg, and \vmicro\ until there
                was no significant dependence of abundances with E.P. and
                RW, and until agreement was attained between
                abundances derived from neutral and ionized species,
                led to the final model parameter choices given in
               the bottom panel.}
                
     \label{fig3}
\end{figure}

To determine the initial model atmosphere parameters, 
we applied the same methods and equations as described in Paper I. 
A detailed description is given in that paper; a brief 
summary is given here.
Photometric effective temperatures were calculated from $(B-V)_0$, and
$(V-K)_0$ colors (Table~\ref{tab2}) using the metallicity-dependent 
\teff-color calibrations given by \cite{ram05} for giant stars.
We also derived LDR temperatures directly from the spectra.
The LDR method was developed by \cite{gra91} and refined by several 
authors \citep[e.g.] {str00,gra01,bia07a,bia07b}. 
We measured the line depths of many lines in the
6190$-$6280 \AA\ spectral region, and formed LDR's for 12 line
pairs out of the 22 that were recommended by \cite{bia07a}.  
Using her polynomial coefficients for 
non-rotationally broadened stars we estimated the individual 
T$_{{\rm eff,LDR}}$ from these LDR's.
Calculated $(B-V)_{0}$, $(V-K)_{0}$\footnote{
The $K$ magnitude is from 2MASS, and \cite{ram05} label the color $(V-K_{2})$.}
and LDR temperatures with their standard deviations are listed in 
Table~\ref{tab6}.
Since $(V-K)_{0}$ colors are nearly independent of 
metallicity, for the stars that have \textit{K} magnitudes we used 
non-weighted averages of T$_{{\rm{eff,({\it V-K})}}}$ 
and T$_{{\rm{eff,LDR}}}$ as initial \teff\ estimates.

We used the following standard equation to compute the initial 
physical gravities of the cluster members. 

\begin{eqnarray}                                           
\logg_{phy} = 0.4~(M_{\rm V\star} + BC - M_{\rm Bol\sun}) + \logg_{\sun}  \nonumber \\  
+4~{\rm log} (\frac{\teff_{\star}}{\teff_{\sun}})+ {\rm log} (\frac{{\it m}_{\star}}{{\it m}_{\sun}}).
\end{eqnarray}

Solar parameters adopted in this equation are: M$_{bol} = 4.75$, log$g$ = 4.44,
and \teff\ = 5777 K.
The polynomial formula and coefficients given in Table~1 of \cite{tor10} 
were used to calculate the temperature-dependent bolometric corrections.
Absolute magnitudes of the target RGs were derived by using the cluster 
distance modulus and reddening (Table~\ref{tab1}).
For the masses of the RGs we used a turn-off mass of 2 
M$_{\odot}$ (see \S\ref{selection}). 
With the guidance of previous studies (\S\ref{previous}), we adopted 
initial metallicities of [M/H] = 0.0 and microturbulent velocities of 
$\xi=1.20$ kms$^{-1}$.

\subsection{Final parameters}\label{final}

\begin{figure}
  \leavevmode
      \epsfxsize=8cm
      \epsfbox{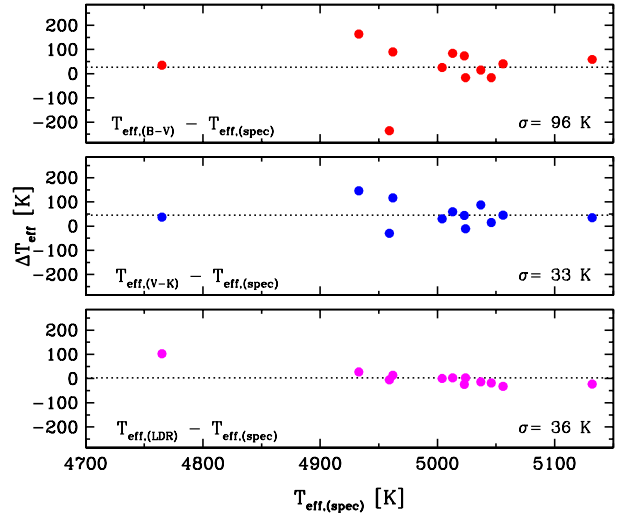}
       \caption{Differences between initial \teff\ estimates 
                and derived spectroscopic temperatures \teff(spec) plotted
                as a function of the T$_{{\rm eff,spec}}$ values.  
                Dotted lines in the panels indicate the mean differences
                and the standard deviations of the differences are given 
                in each panel.
                The top panel contains photometric T$_{{\rm eff,({\it B-V})}}$ 
                values, the middle panel has photometric 
                T$_{{\rm eff,({\it V-K})}}$ values, and the bottom panel has 
                spectroscopic T$_{{\rm eff,LDR}}$ values.}
     \label{fig4}
\end{figure}

\begin{figure}
  \leavevmode
      \epsfxsize=8cm
      \epsfbox{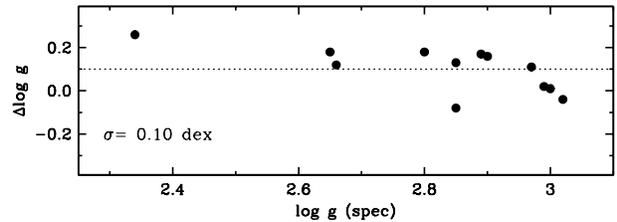}
       \caption{Differences between the physical gravities 
                $\mathrm{log} \ g_{\textit{phy}}$ computed with Eq.~1 and
                the spectroscopic gravities $\mathrm{log} \ g_{\textit{spec}}$
                derived from ionization balances of Fe and Ti, plotted
                as a function of $\mathrm{log} \ g_{\textit{spec}}$.
                A dotted line indicates the mean difference.}
     \label{fig5}
\end{figure}

\begin{table*}
 \centering
 \begin{minipage}{140mm}
  \caption{Photometric and spectroscopic atmospheric parameters.
  \label{tab6}}
  \begin{tabular}{@{}lcccccccr@{}}
  \hline
Star   &  T$_{\textit{eff,(B-V)}}$  &  $T_{\textit{eff,(V-K)}}$   & $T_{\textit{eff,(LDR)}}$   &
$\mathrm{log} \ g_{\textit{,phy}}$  &  $T_{\textit{eff,spec}}$   &  $\mathrm{log} \ g_{\textit{,spec}}$  & $\xi_{\textit{spec}}$ &
$[M/H]$    \\
  &  (K) &  (K)  & (K)   &   (cm s$^{-2}$)  & (K)    & (cm s$^{-2}$)  & (km s$^{-1}$)  &     \\                           
\hline
\multicolumn{9}{c}{NGC~6940}    \\
MMU~28  	&	5008	&5013&	$	5027	\pm33	$	&	3.06	&	5024	&	2.89	&	1.03	&	$	-0.05	$	\\
MMU~30   	&	4724	&4929&	$	4954	\pm39	$	&	2.77	&	4959	&	2.85	&	1.32	&	$	-0.06	$	\\
MMU~60	   &	5030	&	5061	&	$	5028	\pm19	$	&	3.08	&	5046	&	2.97	&	0.97	&	$	-0.02	$	\\
MMU~69	   &	5030	&	5034	&	$	5004	\pm29	$	&	3.06	&	5004	&	2.90	&	1.05	&	$	-0.03	$	\\
MMU~87	   &	5097	&	5067	&	$	4999	\pm43	$	&	2.98	&	5023	&	2.85	&	1.07	&	$	0.03	$	\\
MMU~101	&	5052	&	5125	&	$	5023	\pm22	$	&	2.98	&	5037	&	3.02	&	1.16	&	$	0.01	$	\\
MMU~105	&	4800	&	4802&	$	4867	\pm36	$	&	2.60	&	4765	&	2.34	&	1.35	&	$	-0.15	$	\\
MMU~108	&	5191	&	5167	&	$	5109	\pm17	$	&	2.98	&	5132	&	2.8	&	1.28	&	$	-0.16	$	\\
MMU~132	&	5052	&	5079&	$	4976	\pm37	$	&	2.83	&	4962	&	2.65	&	1.29	&	$	-0.12	$	\\
MMU~138	&	5097	&	5101	&	$	5024	\pm30	$	&	3.01	&	5056	&	3.00	&	1.10	&	$	0.00	$	\\
MMU~139	&	5097	&	5072	&	$	5016	\pm28	$	&	3.01	&	5013	&	2.99	&	1.10	&	$	0.01	$	\\
MMU~152	&	5097	&	5079&	$	4960	\pm12	$	&	2.78	&	4933	&	2.66	&	1.36	&	$	-0.16	$	\\
\multicolumn{9}{c}{Hyades}    \\
$\delta$~Tau           &	4872	&	4918	&	$	4962\pm	33$	&	2.65	&	4878	&	2.57	&	1.34	&	$-0.03$		\\
$\varepsilon$~Tau	&	4812	&	4868	&	$	4921\pm	30$	&	2.53	&	4870	&	2.67	&	1.46	&	$-0.01$		\\
$\gamma$~Tau	   &	4852	&	4928	&	$	4942\pm	32$	&	2.60	&	4945	&	2.78	&	1.42	&	$-0.03$		\\
$\theta$~Tau	       &	4956	&	4980	&	$	4941\pm	43$	&	2.69	&	4961	&	3.00	&	1.28	&	$0.01$		\\
\hline
\end{tabular}
\end{minipage}
\end{table*}

To obtain the final model atmosphere parameters, we used 
the $EW$s of neutral and ionized Fe and Ti lines, initial model atmosphere 
parameters, cluster distance modules and reddening along with the 
magnitudes of RGs as inputs in the semi-automated model iteration code. 
Stellar atmosphere models with opacity distribution functions 
and no convective overshooting were calculated using 
the grids
of ATLAS9 Model Atmospheres from \cite{kur03}. 
Software developed by Andy McWilliam and Inese Ivans that interpolates the 
grids to achieve the final model atmosphere parameters was used
as a part of the model iterations.
In Figure~\ref{fig3} we plot final line-by-line abundance results for the 
star MMU~101, chosen to illustrate the final outcome of our model 
iteration procedure.
As seen in the figure, abundances of \mbox{Fe\,{\sc i}}, \mbox{Fe\,{\sc ii}},
\mbox{Ti\,{\sc i}} and \mbox{Ti\,{\sc ii}} show no obvious trends with 
$\chi$ and $RW$ values and also no abundance differences between the
neutral and ionized species are observed. 
We list the initial and final 
model atmosphere parameters for our NGC~6940 
and Hyades program stars in Table~\ref{tab6}. 

Figure~\ref{fig4}  shows comparisons of initial photometric and LDR 
effective temperatures with the final spectroscopic temperatures 
T$_{{\rm eff,spec}}$. 
Inspection of this figure indicates agreement among these four \teff\ estimates.
Taking into account the data from all three figure panels, 
the average of the difference between initial photometric and LDR 
temperatures and final spectroscopic \teff's is 
$\langle {\rm T_{\rm eff,initial} - T_{\rm eff,spec}}\rangle$ = 26~K.
Both photometric temperatures T$_{\rm{eff,({\it B-V})}}$ and 
T$_{\rm{eff,({\it V-K})}}$ suffer from uncertainties that arise from 
reddening and distance parameters. 
For two of the RGs the T$_{\rm{eff,({\it B-V})}}$ values deviate 
significantly from T$_{\rm{eff,spec}}$ by $-235$ and $+164$~K. 
They create the apparent large scatter in the top panel of 
Figure~\ref{fig4}.
Even so, the mean of the differences is just 
$\langle {\rm T_{eff,({\it B-V})} - T_{eff,spec}}\rangle$ = 27 $\pm$ 28.
T$_{\rm{eff,({\it V-K})}}$ color tempratures correlate well with spectroscopic ones,
$\langle {\rm T_{eff,({\it V-K})} - T_{eff,spec}}\rangle$ = 48 $\pm$ 12.
All photometric and LDR temperatures are in good agreement 
with the T$_{\rm{eff,spec}}$ except for MMU~105, with
$\langle {\rm T_{eff,LDR} - T_{eff,spec}}\rangle$ = 2.8$\pm$10~K.
A similar exercise for the Hyades also yields good agreement among the
temperature estimates, with the mean offset being $-$1~K ($\sigma$ =  36~K).

In Figure~\ref{fig5} we show the differences between the 
calculated physical gravities (initial) and spectroscopic gravities (final). 
The initial and final gravities agree well within 
the uncertainties introduced by other parameters in Eq.~1.
Defining $\Delta \logg =\logg_{\textit{phy}} - \logg_{\textit{spec}}$
we find $\langle \Delta \logg \rangle$ = 0.10 ($\sigma$ = 0.10).
For the Hyades the agrement is also good, with the mean offset being
$-$0.14 ($\sigma$ = 0.16).

Since the individual standard deviations of the metallicities are about 
0.06 dex, the differences in metallicities of the members stay within 
the uncertainty limits and do not rule out the membership status of the RGs. 
The mean metallicity calculated from both Fe and Ti lines
for NGC~6940 from 12 RGs is 
$\langle {\rm [M/H]} \rangle = -0.06\pm0.07$.

\subsection{Parameter uncertainties\label{paramerror}}

We followed a very similar technique as described in Paper~1 to estimate 
the internal uncertainties for the atmospheric parameters. 
First we selected three RGs, MMU~28, MMU~101 and MMU~105, as the 
representatives for our sample and then applied series of analyses to 
their spectra to 
estimate the probable uncertainties in effective 
temperature, microturbulance velocity and gravity. 

Taking each star in turn, for \teff~we changed the 
temperature by 50~K in each step (while keeping the other parameters fixed) 
until the average difference in abundances between low and high excitation 
potential \mbox{Fe\,{\sc i}} lines exceeded the line-to-line scatter 
$\sigma$ of the abundances in the final derived model atmosphere.
The average uncertainty for \teff\ computed in this manner was $\sim100$~K.
We did similar calculation for the microturbulance, changing \vmicro\ in steps
of  0.1 \kmsec\ and seeing the abundance effects on neutral and ionized 
species of Fe and Ti, led to a $\vmicro$ uncertainty in $\vmicro$ of 
$\sim0.2$ \kmsec.
Finally, changing the surface gravity in stops of 0.05 dex and computing 
abundance differences between neutral and ionized species of these two
elements led to a \logg\ uncercertainty of $\sim0.20$ dex.
Unfortunately we could not investigate the external uncertainties for our 
sample since this is first detailed spectroscopic study of the NGC~6940 
RGs in the literature.
 
The standard deviation ($\sim0.1$ dex) of the differences 
between the physical and spectroscopic \logg\ values are shown in Figure~\ref{fig5}. 
Since the given $\sigma$ is smaller than what was found from the internal 
uncertainty determinations, we did not include it to our overall uncertainties.
The effect of the uncertainties on the elemental abundances is given in Table~\ref{tab7}.
We only summarize the results from the analysis of MMU~101, which in turn 
represents the overall sample. 
Since the observed spectrum at 8000~\AA\ used for the \ciso\ 
determination of MMU~101 (\ciso\ $>$ 25) is relatively low, we estimated 
the \ciso\ uncertainties from the spectrum of MMU~152. 
$^{12}$CN and $^{13}$CN molecular lines are not really sensitive to
the changes in model atmosphere parameters as it is seen also in 
Table~\ref{tab7}. 

  \begin{table}
 \centering
 \begin{minipage}{90mm}
  \caption{Sensitivity ($\sigma$) of derived abundances to the
   model atmosphere changes within uncertainty limits for the star MMU~101.}
  \label{tab7}
  \begin{tabular}{@{}lccc@{}}
  \hline
Species &  $\Delta$\teff (K) &  $\Delta$\logg  & $\Delta$\vmicro (kms$^{-1}$) \\
 & $-$100 $/$ $+$ 100  &  $-$0.20 $/$ $+$0.20  &  $-$0.2 $/$ $+$0.2   \\
\hline
\mbox{Li\,{\sc i}}	&	$+$	0.10	$/$	$-$	0.10	&	~~	0.00	$/$	$-$	0.05	&	$-$	0.05	$/$	~~	0.00	\\
C     	&	$+$	0.02	$/$	$-$	0.05	&	$+$	0.03	$/$	$-$	0.02	&	$+$	0.01	$/$	~~	0.00	\\
N     	&	$+$	0.10	$/$	$-$	0.10	&	$+$	0.02	$/$	$-$	0.01	&	~~	0.00	$/$	~~	0.00	\\
O     	&	$+$	0.05	$/$	$-$	0.05	&	$+$	0.12	$/$	$-$	0.10	&	~~	0.00	$/$	~~	0.00	\\
\mbox{Na\,{\sc i}}	&	$+$	0.09	$/$	$-$	0.06	&	$-$	0.01	$/$	$+$	0.01	&	$-$	0.02	$/$	$+$	0.06	\\
\mbox{Mg\,{\sc i}}	&	$+$	0.05	$/$	$-$	0.09	&	$-$	0.02	$/$	$+$	0.02	&	$-$	0.08	$/$	$+$	0.05	\\
\mbox{Al\,{\sc i}}	&	$+$	0.06	$/$	$-$	0.05	&	~~	0.00	$/$	~~	0.00	&	~~	0.00	$/$	$+$	0.03	\\
\mbox{Si\,{\sc i}}	&	$-$	0.02	$/$	$+$	0.01	&	$+$	0.02	$/$	$-$	0.04	&	$-$	0.04	$/$	$+$	0.04	\\
\mbox{Ca\,{\sc i}}	&	$+$	0.09	$/$	$-$	0.09	&	$-$	0.03	$/$	$+$	0.03	&	$-$	0.09	$/$	$+$	0.09	\\
\mbox{Sc\,{\sc ii}}	&	$-$	0.01	$/$	~~	0.00	&	$+$	0.07	$/$	$-$	0.10	&	$-$	0.08	$/$	$+$	0.07	\\
\mbox{Ti\,{\sc i}}	&	$+$	0.13	$/$	$-$	0.13	&	~~	0.00	$/$	$+$	0.01	&	$-$	0.04	$/$	$+$	0.03	\\
\mbox{Ti\,{\sc ii}}	&	$-$	0.02	$/$	$+$	0.01	&	$+$	0.07	$/$	$-$	0.10	&	$-$	0.12	$/$	$+$	0.11	\\
\mbox{V\,{\sc i}}	&	$+$	0.15	$/$	$-$	0.15	&	$+$	0.01	$/$	~~	0.00	&	$-$	0.02	$/$	$+$	0.02	\\
\mbox{Cr\,{\sc i}}	&	$+$	0.10	$/$	$-$	0.10	&	$-$	0.01	$/$	$+$	0.02	&	$-$	0.06	$/$	$+$	0.06	\\
\mbox{Cr\,{\sc ii}}	&	$-$	0.06	$/$	$+$	0.05	&	$+$	0.07	$/$	$-$	0.11	&	$-$	0.08	$/$	$+$	0.07	\\
\mbox{Mn\,{\sc i}}	&	$+$	0.12	$/$	$-$	0.10	&	~~	0.00	$/$	$+$	0.02	&	$-$	0.03	$/$	$+$	0.07	\\
\mbox{Fe\,{\sc i}}	&	$+$	0.06	$/$	$-$	0.07	&	~~	0.00	$/$	~~	0.00	&	$-$	0.09	$/$	$+$	0.09	\\
\mbox{Fe\,{\sc ii}}	&	$-$	0.08	$/$	$+$	0.07	&	$+$	0.08	$/$	$-$	0.13	&	$-$	0.07	$/$	$+$	0.06	\\
\mbox{Co\,{\sc i}}	&	$+$	0.08	$/$	$-$	0.09	&	$+$	0.03	$/$	$-$	0.02	&	$-$	0.01	$/$	$+$	0.01	\\
\mbox{Ni\,{\sc i}}	&	$+$	0.04	$/$	$-$	0.05	&	$+$	0.02	$/$	$-$	0.03	&	$-$	0.08	$/$	$+$	0.07	\\
\mbox{Cu\,{\sc i}}	&	$+$	0.05	$/$	$-$	0.08	&	$+$	0.05	$/$	$-$	0.02	&	$-$	0.05	$/$	$+$	0.07	\\
\mbox{Zn\,{\sc i}}	&	$+$	0.05	$/$	$-$	0.05	&	~~	0.00	$/$	$-$	0.10	&	~~	0.00	$/$	$-$	0.03	\\
\mbox{Y\,{\sc ii}}	&	$+$	0.05	$/$	$-$	0.04	&	$+$	0.07	$/$	$-$	0.07	&	$-$	0.06	$/$	$+$	0.06	\\
\mbox{La\,{\sc ii}}	&	$+$	0.02	$/$	$-$	0.04	&	$+$	0.07	$/$	$-$	0.10	&	$-$	0.02	$/$	~~	0.00	\\
\mbox{Nd\,{\sc ii}}	&	$+$	0.05	$/$	$-$	0.05	&	$+$	0.10	$/$	$-$	0.08	&	$-$	0.07	$/$	$+$	0.05	\\
\mbox{Eu\,{\sc ii}}	&	$-$	0.02	$/$	$-$	0.02	&	$+$	0.06	$/$	$-$	0.13	&	$-$	0.03	$/$	~~	0.00	\\
$^{12}C/^{13}C$       &      $-$   1 $/$ $+$  1  &        0 $/$ $+$  1   &      0 $/$ 0  \\
\hline
\end{tabular}
\end{minipage}
\end{table}

\section{Derived Abundances}\label{abunds}

\begin{table}
 \centering
 \begin{center}
 \begin{minipage}{70mm}
  \caption{Solar abundances. Species noted with (*) are updated abundances, 
  the others are from Paper~1.}
  \label{tab8}
  \begin{tabular}{@{}lcc@{}}
  \hline
Species & log $\epsilon_{\odot}$  & log $\epsilon_{\odot}$     \\ 
 &     &  \citep{asp09}       \\                             
\hline
\mbox{Li\,{\sc i}}       &   1.05$\pm(0.05)$			       &   1.05$\pm0.10$     \\
C         &   8.43$\pm(0.05)$  			     &   8.43$\pm0.05$      \\
N$^{*}$      &   7.99$\pm(0.05)$ 			       &   7.83$\pm0.05$       \\
O         &   8.69$\pm(0.05)$       			 &   8.69$\pm0.05$       \\  
\mbox{Na\,{\sc i}}       &   6.34$\pm(0.10)$				        &   6.24$\pm0.04$       \\
\mbox{Mg\,{\sc i}}     &   7.63$\pm (0.16)$                      &   7.6$\pm0.04$       \\
\mbox{Al\,{\sc i}}        &   6.33$\pm (0.18)$                     &   6.45$\pm0.03$        \\
\mbox{Si\,{\sc i}}$^{*}$      &   7.55$\pm0.06$    &   7.51$\pm0.03$       \\
\mbox{Ca\,{\sc i}}$^{*}$     &   6.32$\pm0.06$    &   6.34$\pm0.04$        \\
\mbox{Sc\,{\sc ii}}     &                                     &   3.15$\pm0.04$       \\
\mbox{Ti\,{\sc i}}$^{*}$      &  4.87$\pm0.06$     &    4.95$\pm0.05$    \\
\mbox{Ti\,{\sc ii}}$^{*}$     &   4.95$\pm0.05$    &                            \\
\mbox{V\,{\sc i}}$^{*}$         &         3.87$\pm0.03$                             &    3.93$\pm0.08$    \\
\mbox{Cr\,{\sc i}}$^{*}$     &    5.60$\pm0.05$    &   5.64$\pm0.04$     \\
\mbox{Cr\,{\sc i}}$^{*}$     &   5.70$\pm0.07$      &                           \\
\mbox{Mn\,{\sc i}}     &   5.41$\pm(0.06)$                     &    5.43$\pm0.04$    \\
\mbox{Fe\,{\sc i}}$^{*}$     &    7.42$\pm0.06$     &    7.50$\pm0.04$    \\
\mbox{Fe\,{\sc ii}}$^{*}$    &     7.41$\pm0.05$    &                             \\
\mbox{Co\,{\sc i}}$^{*}$    &    4.92$\pm0.07$      &    4.99$\pm0.07$    \\
\mbox{Ni\,{\sc i}}$^{*}$     &   6.26$\pm0.07$     &     6.22$\pm0.04$    \\
\mbox{Cu\,{\sc i}}    &    4.23$\pm(0.10)$                     &    4.19$\pm0.04$    \\
\mbox{Zn\,{\sc i}}      &   4.51$\pm(0.05)$  					      &     4.56$\pm0.05$    \\
\mbox{Y\,{\sc ii}}       &    2.23$\pm(0.04)$                       &    2.21$\pm0.05$     \\
\mbox{La\,{\sc ii}}      &   1.15$\pm(0.06)$                       &     1.10$\pm0.04$   \\
\mbox{Nd\,{\sc ii}}      &   1.39$\pm(0.05)$  				       &    1.42$\pm0.04$   \\
\mbox{Eu\,{\sc ii}}      &   0.54$\pm(0.08)$                    &    0.52$\pm0.04$   \\
\hline
\end{tabular}
\end{minipage}
\end{center}
\end{table}

\begin{table*}
 \centering
 \begin{minipage}{250mm}
 \caption{Elemental abundances of individual stars and average abundances ($\langle$[X/Fe]$\rangle$) for NGC~6940.\label{tab9}}
  \begin{tabular}{@{}lrrrrrrrrrrrrcc@{}}
  \hline
Species  & \multicolumn{12}{c}{MMU} \\
$\rm{[X/Fe]}$& 28  & 30  & 60  &  69  &  87  &  101  &  105  &  108  &  132  &  138  & 139 & 152 & $\langle$[X/Fe]$\rangle$ & $\sigma$\\ 
\hline
\rm{C}	&	$	-0.29	$	&	$	-0.25	$	&	$	-0.38	$	&	$	-0.31	$	&	$	-0.29	$	&	$	-0.33	$	&	$	-0.34	$	&	$	-0.26	$	&	$	-0.39	$	&	$	-0.22	$	&	$	-0.29	$	&	$	-0.49	$	&	$	-0.32	$&$	0.07	$	\\
\rm{N}		&	$	0.19	$	&	$	0.19	$	&	$	0.23	$	&	$	0.26	$	&	$	0.19	$	&	$	0.25	$	&	$	0.19	$	&	$	0.37	$	&	$	0.21	$	&	$	0.24	$	&	$	0.16	$	&	$	0.34	$	&	$	~~ 0.23$&$		0.06	$	\\
\mbox{O\,{\sc i}}  	&	$	-0.15	$	&	$	-0.11	$	&	$	-0.25	$	&	$	-0.14	$	&	$	-0.12	$	&	$	-0.12	$	&	$	-0.20	$	&	$	-0.14	$	&	$	-0.27	$	&	$	-0.06	$	&	$	-0.12	$	&	$	-0.13	$	&	$	-0.15	$&$	0.06	$	\\
\mbox{Na\,{\sc i}}	&	$	0.06	$	&	$	0.06	$	&	$	0.01	$	&	$	0.07	$	&	$	-0.05	$	&	$	0.06	$	&	$	0.18	$	&	$	0.31	$	&	$	0.19	$	&	$	0.00	$	&	$	0.04	$	&	$	0.38	$	&	$	~~ 0.11$&$	0.13	$	\\
\mbox{Mg\,{\sc i}}	&	$	-0.04	$	&	$	-0.08	$	&	$	-0.05	$	&	$	-0.04	$	&	$	-0.08	$	&	$	-0.02	$	&	$	0.01	$	&	$	0.09	$	&	$	0.01	$	&	$	-0.03	$	&	$	-0.05	$	&	$	0.05	$	&	$	-0.02	$&$	0.05	$	\\
\mbox{Al\,{\sc i}}	&	$	-0.02	$	&	$	-0.04	$	&	$	-0.07	$	&	$	0.03	$	&	$	-0.02	$	&	$	-0.03	$	&	$	0.04	$	&	$	0.16	$	&	$	0.05	$	&	$	0.00	$	&	$	0.00	$	&	$	0.12	$	&	$	~~ 0.02	$&$	0.07	$	\\
\mbox{Si\,{\sc i}}	&	$	0.10	$	&	$	0.13	$	&	$	0.00	$	&	$	0.11	$	&	$	0.07	$	&	$	0.12	$	&	$	0.17	$	&	$	0.13	$	&	$	0.20	$	&	$	0.05	$	&	$	0.13	$	&	$	0.21	$	&	$	~~ 0.12$&$	0.06	$	\\
\mbox{Ca\,{\sc i}}	&	$	0.09	$	&	$	0.03	$	&	$	0.07	$	&	$	0.12	$	&	$	0.05	$	&	$	0.04	$	&	$	0.11	$	&	$	0.15	$	&	$	0.12	$	&	$	0.01	$	&	$	0.08	$	&	$	0.09	$	&	$	~~ 0.08	$&$	0.04	$	\\
\mbox{Sc\,{\sc ii}}	&	$	-0.05	$	&	$	-0.03	$	&	$	-0.04	$	&	$	0.00	$	&	$	-0.07	$	&	$	0.04	$	&	$	-0.11	$	&	$	-0.08	$	&	$	-0.07	$	&	$	-0.03	$	&	$	-0.02	$	&	$	-0.04	$	&	$	-0.04	$&$	0.04	$	\\
\mbox{Ti\,{\sc i}}	&	$	-0.04	$	&	$	-0.06	$	&	$	-0.05	$	&	$	-0.06	$	&	$	-0.09	$	&	$	-0.06	$	&	$	-0.06	$	&	$	0.07	$	&	$	-0.01	$	&	$	0.01	$	&	$	-0.07	$	&	$	-0.01	$	&	$	-0.04	$&$	0.04	$	\\
\mbox{Ti\,{\sc ii}}	&	$	-0.03	$	&	$	0.04	$	&	$	-0.03	$	&	$	-0.03	$	&	$	-0.04	$	&	$	0.01	$	&	$	-0.02	$	&	$	0.03	$	&	$	0.04	$	&	$	0.01	$	&	$	0.08	$	&	$	-0.03	$	&	$	~~ 0.00	$&$	0.04	$	\\
\mbox{V\,{\sc i}}	&	$	-0.04	$	&	$	-0.04	$	&	$	-0.09	$	&	$	-0.06	$	&	$	-0.12	$	&	$	-0.04	$	&	$	-0.06	$	&	$	0.04	$	&	$	-0.02	$	&	$	-0.01	$	&	$	-0.08	$	&	$	0.00	$	&	$	-0.04	$&$	0.04	$	\\
\mbox{Cr\,{\sc i}}	&	$	0.04	$	&	$	0.02	$	&	$	0.03	$	&	$	0.07	$	&	$	0.01	$	&	$	0.02	$	&	$	0.05	$	&	$	0.12	$	&	$	0.05	$	&	$	0.08	$	&	$	-0.01	$	&	$	0.06	$	&	$	~~ 0.04	$&$	0.03	$	\\
\mbox{Cr\,{\sc ii}}	&	$	0.06	$	&	$	0.12	$	&	$	0.03	$	&	$	0.00	$	&	$	0.06	$	&	$	0.05	$	&	$	0.10	$	&	$	0.01	$	&	$	0.20	$	&	$	-0.01	$	&	$	0.09	$	&	$	0.04	$	&	$	~~ 0.06	$&$	0.06	$	\\
\mbox{Mn\,{\sc i}}	&	$	-0.11	$	&	$	-0.12	$	&	$	-0.13	$	&	$	-0.13	$	&	$	-0.16	$	&	$	-0.13	$	&	$	-0.20	$	&	$	-0.09	$	&	$	-0.06	$	&	$	-0.11	$	&	$	-0.15	$	&	$	-0.06	$	&	$	-0.12	$&$	0.04	$	\\
$[${\mbox{Fe\,{\sc i}}/H}$]$	&	$	0.05	$	&	$	0.09	$	&	$	0.12	$	&	$	0.05	$	&	$	0.14	$	&	$	0.09	$	&	$	-0.02	$	&	$	-0.07	$	&	$	0.07	$	&	$	0.08	$	&	$	0.09	$	&	$	0.02	$	&	$	~~ 0.06$&$		0.06	$	\\
$[${\mbox{Fe\,{\sc ii}}/H}$]$	&	$	0.02	$	&	$	0.05	$	&	$	0.10	$	&	$	0.03	$	&	$	0.12	$	&	$	0.07	$	&	$	-0.06	$	&	$	-0.13	$	&	$	0.01	$	&	$	0.04	$	&	$	0.07	$	&	$	-0.04	$	&	$	~~ 0.02	$&$	0.07	$	\\
\mbox{Co\,{\sc i}}	&	$	-0.12	$	&	$	-0.08	$	&	$	-0.15	$	&	$	-0.13	$	&	$	-0.17	$	&	$	-0.09	$	&	$	-0.12	$	&	$	-0.09	$	&	$	-0.09	$	&	$	-0.10	$	&	$	-0.11	$	&	$	-0.05	$	&	$	-0.11	$&$	0.03	$	\\
\mbox{Ni\,{\sc i}}	&	$	-0.02	$	&	$	-0.02	$	&	$	-0.04	$	&	$	-0.02	$	&	$	-0.03	$	&	$	-0.01	$	&	$	0.01	$	&	$	-0.03	$	&	$	-0.01	$	&	$	-0.02	$	&	$	0.03	$	&	$	0.01	$	&	$	-0.01	$&$	0.02	$	\\
\mbox{Cu\,{\sc i}}&	$	-0.10	$	&	$	-0.03	$	&	$	-0.14	$	&	$	-0.13	$	&	$	-0.16	$	&	$	-0.10	$	&	$	-0.16	$	&	$	-0.11	$	&	$	-0.05	$	&	$	-0.08	$	&	$	-0.07	$	&	$	0.02	$	&	$	-0.09	$&$	0.05	$	\\
\mbox{Zn\,{\sc i}}	&	$	0.06	$	&	$	0.07	$	&	$	0.09	$	&	$	0.08	$	&	$	0.00	$	&	$	0.08	$	&	$	0.17	$	&	$	0.04	$	&	$	0.09	$	&	$	0.04	$	&	$	0.18	$	&	$	0.03	$	&	$	~~ 0.08	$&$	0.05	$	\\
\mbox{Y\,{\sc ii}}	&	$	0.00	$	&	$	0.02	$	&	$	-0.05	$	&	$	-0.01	$	&	$	-0.06	$	&	$	0.02	$	&	$	-0.05	$	&	$	0.07	$	&	$	-0.07	$	&	$	-0.04	$	&	$	0.03	$	&	$	0.00	$	&	$	-0.01	$&$	0.04	$	\\
\mbox{La\,{\sc ii}}&	$	0.04	$	&	$	0.10	$	&	$	-0.03	$	&	$	0.08	$	&	$	0.03	$	&	$	0.14	$	&	$	0.00	$	&	$	0.06	$	&	$	0.04	$	&	$	0.09	$	&	$	0.06	$	&	$	0.05	$	&	$	~~ 0.05	$&$	0.05	$	\\
\mbox{Nd\,{\sc ii}}	&	$	0.19	$	&	$	0.24	$	&	$	0.15	$	&	$	0.15	$	&	$	0.22	$	&	$	0.26	$	&	$	0.15	$	&	$	0.12	$	&	$	0.16	$	&	$	0.15	$	&	$	0.17	$	&	$	0.23	$	&	$	~~ 0.18	$&$	0.04	$	\\
\mbox{Eu\,{\sc ii}}	&	$	-0.06	$	&	$	0.03	$	&	$	-0.01	$	&	$	0.01	$	&	$	-0.04	$	&	$	0.03	$	&	$	0.01	$	&	$	0.01	$	&	$	-0.02	$	&	$	0.07	$	&	$	0.02	$	&	$	0.01	$	&	$	~~ 0.00	$&$	0.03	$	\\
 \hline
$[${\mbox{N/C}}$]$  &  $0.48$  & $0.44$  &  $0.61$   &   $0.57$  & $0.48$ & $0.58$  &  $0.53$  & $0.63$ &  $0.61$  &  $0.46$  &  $0.45$  &  $0.83$  \\
$\rm{log~\epsilon(Li)}$  &	$ 1.05$     &	$	<0	$       &	$	1.23	$     &	$	1.29	$     &	$	0.56	$     &	$	0.64	$     &	$	<0.1 $     &	$	<0	    $  &	$    0.33	$      &	$	1.05	$ &	$	0.86	$  &	$    <0	$ \\
\hline
\end{tabular}
\end{minipage}
\end{table*}

\begin{table}
 \centering
 \begin{minipage}{90mm}
 \caption{Elemental abundances of individual stars and average abundances ($<$[X/Fe]$>$) for Hyades.
 \label{tab10}}
  \begin{tabular}{@{}lrrrrcc@{}}
      \hline
  Species  & 	$\delta$~Tau &	$\varepsilon$~Tau & $\gamma$~Tau	 & $\theta$~Tau	& $<$[X/Fe]$>$ & $\sigma$ \\
$\rm{[X/Fe]}$&&&&&&\\
    \hline
\rm{C}	&	$	-0.49	$	&	$	-0.45	$	&	$	-0.42	$	&	$	-0.44	$	&	$	-0.45	$	&	$	0.03	$\\
\rm{N}	&	$	0.33	$	&	$	0.37	$	&	$	0.44	$	&	$	0.36	$	&	$	~~ 0.38	$	&	$	0.04	$\\
\mbox{O\,{\sc i}}  	&	$	-0.28	$	&	$	-0.19	$	&	$	-0.20	$	&	$	-0.15	$	&	$	-0.20	$	&	$	0.06	$\\
\mbox{Na\,{\sc i}}	&	$	0.21	$	&	$	0.18	$	&	$	0.23	$	&	$	0.14	$	&	$	~~ 0.19	$	&	$	0.04	$\\
\mbox{Mg\,{\sc i}}	&	$	-0.01	$	&	$	0.00	$	&	$	-0.04	$	&	$	-0.10	$	&	$	-0.04	$	&	$	0.04	$\\
\mbox{Al\,{\sc i}}	&	$	0.04	$	&	$	0.09	$	&	$	0.07	$	&	$	0.01	$	&	$	~~ 0.05	$	&	$	0.03	$\\
\mbox{Si\,{\sc i}}	&	$	0.27	$	&	$	0.26	$	&	$	0.22	$	&	$	0.21	$	&	$	~~ 0.24	$	&	$	0.03	$\\
\mbox{Ca\,{\sc i}}	&	$	0.08	$	&	$	0.03	$	&	$	0.06	$	&	$	0.02	$	&	$	~~ 0.05	$	&	$	0.03	$\\
\mbox{Sc\,{\sc ii}}	&	$	-0.02	$	&	$	0.04	$	&	$	-0.01	$	&	$	0.09	$	&	$	~~ 0.02	$	&	$	0.05	$\\
\mbox{Ti\,{\sc i}}	&	$	-0.09	$	&	$	-0.08	$	&	$	-0.03	$	&	$	-0.01	$	&	$	-0.05	$	&	$	0.04	$\\
\mbox{Ti\,{\sc ii}}	&	$	-0.05	$	&	$	-0.01	$	&	$	0.05	$	&	$	0.05	$	&	$	~~ 0.01	$	&	$	0.05	$\\
\mbox{V\,{\sc i}}	&	$	-0.09	$	&	$	-0.02	$	&	$	0.01	$	&	$	-0.01	$	&	$	-0.03	$	&	$	0.04	$\\
\mbox{Cr\,{\sc i}}	&	$	0.06	$	&	$	0.05	$	&	$	0.04	$	&	$	0.03	$	&	$	~~ 0.04	$	&	$	0.01	$\\
\mbox{Cr\,{\sc ii}}	&	$	0.12	$	&	$	0.22	$	&	$	0.14	$	&	$	0.12	$	&	$	~~ 0.15	$	&	$	0.05	$\\
\mbox{Mn\,{\sc i}}	&	$	-0.14	$	&	$	-0.14	$	&	$	-0.13	$	&	$	-0.18	$	&	$	-0.15	$	&	$	0.02	$\\
$[${\mbox{Fe\,{\sc i}}/H}$]$	&	$	0.08	$	&	$	0.14	$	&	$	0.12	$	&	$	0.15	$	&	$	~~ 0.12	$	&	$	0.03	$\\
$[${\mbox{Fe\,{\sc ii}}/H}$]$	&	$	0.07	$	&	$	0.11	$	&	$	0.12	$	&	$	0.16	$	&	$	~~ 0.12	$	&	$	0.04	$\\
\mbox{Co\,{\sc i}}	&	$	-0.09	$	&	$	-0.06	$	&	$	-0.04	$	&	$	-0.04	$	&	$	-0.06	$	&	$	0.02	$\\
\mbox{Ni\,{\sc i}}	&	$	0.01	$	&	$	0.06	$	&	$	0.03	$	&	$	0.07	$	&	$	~~ 0.04	$	&	$	0.03	$\\
\mbox{Cu\,{\sc i}}	&	$	-0.07	$	&	$	-0.05	$	&	$	-0.04	$	&	$	-0.07	$	&	$	-0.06	$	&	$	0.01	$\\
\mbox{Zn\,{\sc i}}	&	$	0.08	$	&	$	-0.04	$	&	$	0.00	$	&	$	0.00	$	&	$	~~ 0.01	$	&	$	0.05	$\\
\mbox{Y\,{\sc ii}}	&	$	-0.14	$	&	$	-0.09	$	&	$	-0.17	$	&	$	-0.17	$	&	$	-0.14	$	&	$	0.04	$\\
\mbox{La\,{\sc ii}}	&	$	-0.12	$	&	$	-0.07	$	&	$	-0.09	$	&	$	-0.02	$	&	$	-0.07	$	&	$	0.04	$\\
\mbox{Nd\,{\sc ii}}	&	$	0.07	$	&	$	0.16	$	&	$	0.15	$	&	$	0.11	$	&	$	~~ 0.12	$	&	$0.04	$\\
\mbox{Eu\,{\sc ii}}	&	$	-0.14	$	&	$	-0.08	$	&	$	-0.11	$	&	$	-0.03	$	&	$	-0.09	$	&	$	0.05	$\\
 \hline
$[${\mbox{N/C}}$]$  &  $0.82$  & $0.83$  &  $0.86$   &  $0.80$     \\
$\rm{log~\epsilon(Li)}$  &	$ 0.87$     &	$	0.54	$       &	$	0.97	$     &	$	1.11     $ \\
 \hline
\end{tabular}
\end{minipage}
\end{table}

As discussed in \S\ref{lines} we derived elemental abundances 
of elements from $EW$ measurements and from synthetic spectrum calculations 
for lines that are blended or have hyperfine and isotopic splitting.
In order to find the differential abundances relative to the Sun, we made 
use of the integrated solar flux atlas of \cite{kur84} and derived 
the solar abundances for the same line list used for the targets. 
We assumed \teff ~$=$~5777 K, log $g=4.44$ cgs, 
$\xi=0.85$ km$^{-1}$ and calculated the solar model atmosphere by using 
\cite{kur03} grids. 
As described in \S\ref{linelists}, our line list has been slightly updated 
for the species that we determine abundances from $EW$ measurements. 
Solar abundances for the updated lines are listed in 
Table~\ref{tab8} along with the ones derived by \cite{asp09}. 
For other
lines that require synthetic spectrum matching technique, we adopted 
the abundances from Table~9 of Paper~1.

In this study, we used new transition probabilities for \mbox{Co\,{\sc i}} 
\citep{law15} and \mbox{V\,{\sc i}} \citep{law14}, 
and so the solar Co and V 
abundances were re-calculated (Table~\ref{tab8}).
We again relied on reverse solar analyses to derive the transition 
probabilities for \mbox{Sc\,{\sc ii}} lines that have no recent lab data.
In Table~\ref{tab9} and Table~\ref{tab10} we list the derived abundances 
of all species [X/Fe] for individual RGs and the mean cluster 
abundances ($\langle$[X/Fe]$\rangle$) of NGC~6940 and Hyades.

In total, we derived the abundances of 26 species of 23 elements. 
We will comment on a few of the computational issues and
discuss the results by grouping the elements 
in the following nuclesynthetic groups;
$\alpha$ (Mg, Si, Ca); light odd-Z (Na, Al); Fe-group
(Sc, Ti, V, Cr, Mn, Fe, Co, Ni, Cu, Zn); $n$-capture (Y, La, Nd, Eu);
and $p$-capture (Li, C, N, O). 
Figure~\ref{fig6} shows the abundances of all elements for the 12 RGs 
of NGC~6940 as listed in Table~\ref{tab9}. 
The mean abundances of most species scattered around solar 
($0.01$ to $0.1$ dex).

\subsection{$\alpha$, Odd-Z, Fe-group and Neutron-capture 
elements\label{alpha}}

\textbf{\textit{$\alpha$ and Odd-Z elements}}: 
We analyzed the $\alpha$ elements Mg, Si and Ca.
Abundances of Si and Ca were derived from 
$EW$ measurements of their neutral species.
The cluster $\langle$[Si/Fe]$\rangle$ is slightly over solar while
$\langle$[Ca/Fe]$\rangle$ is about solar (Table~\ref{tab9}).
Mg abundances were determined from synthetic spectrum fitting to
the Mg lines at 5528.4~\AA\ and 5711.1~\AA; 
both lines are slightly contaminated by C$_{2}$ and CN molecular absorptions. 
$\langle$[Mg/Fe]$\rangle$ is solar (Table~\ref{tab9}), as is the cluster 
mean of all three alphas: \\
$\langle$[$\alpha$/Fe]$\rangle$ $\equiv$ \begin{scriptsize}$\dfrac{1}{3}$\end{scriptsize}
([Mg/Fe]+[Ca/Fe]+[Si/Fe]) = $0.06\pm0.07$.

Abundance determination of two odd-Z light elements Na and 
Al were made via synthetic spectra.
\mbox{Na\,{\sc i}} lines at 5682.6, 5688.2, 6154.2 and 
6160.8~\AA\ were analyzed; the Na D lines were far too strong to be useful. 
According to \cite{tak03} the Na lines we analyzed 
are mildly affected by non-LTE in disk stars. Their calculations indicate that 
non-LTE corrections are quite small and about $-$0.1 and $-$0.05 dex for 
5682/5688 and 6154/6160 absorption lines, respectively. 
These non-LTE correction factors are small enough that
we chose not to apply them to our results.

\begin{figure}
  \leavevmode
      \epsfxsize=8.2cm
      \epsfbox{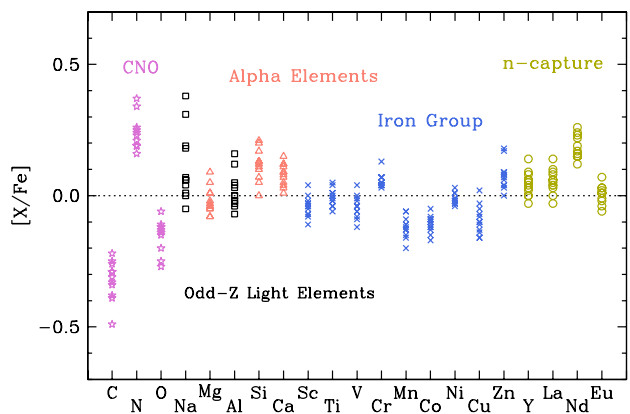}
       \caption{[X/Fe] values of the species studied for the entire RG sample. 
                Dotted line represents the solar values. Abundances of 
                CNO, odd-Z, alpha, iron-group and n-capture elements are 
                shown by stars, squares, triangles, crosses and empty 
                circles, respectively.}
     \label{fig6}
\end{figure}

The mean Na abundance for the cluster is 
$\langle$[Na/Fe]$\rangle$ = $0.11\pm0.13$. 
Na abundances in MMU~108 and MMU~152, depart from the
solar abundance by about 0.3 dex and 0.4 dex, respectively.  
Possible reasons for the differences in Na abundances
will be discussed in \S\ref{discuss}.
Al abundances of the stars were determined from four \mbox{Al\,{\sc i}} 
lines; 6696.1, 6696.7, 7835.3 and 7836.1~\AA; all are blended by atomic 
lines and thus needed synthetic spectrum treatment.
The mean cluster abundance of this element is solar:
$\langle$[Al/Fe]$\rangle$ = $0.02\pm0.07$. 
The Al abundances of MMU~108 and MMU~152 also 
depart from the solar abundance by about $\sim$0.14 dex.
Deviation of the Na and Al abundances of these RGs from the mean
can be seen in Figure~\ref{fig6}.

\textbf{\textit{Fe-group elements}}:
Abundances of Ti, Cr, and Ni were treated to single-line
$EW$ analyses, all of which yielded $\langle$[X/Fe]$\rangle$ $\simeq$ 0
(Table~\ref{tab9}, Figure~\ref{fig6}).
Ti abundances from \mbox{Ti\,{\sc i}} and \mbox{Ti\,{\sc ii}} lines have
already been considered as part of model atmosphere determinations 
(\S\ref{final}).
We used 7$-$10 \mbox{Ti\,{\sc i}} and 2$-$4 \mbox{Ti\,{\sc ii}} lines
for each star. 
For Cr, 6$-$13 \mbox{Cr\,{\sc i}} lines and 2$-$3 \mbox{Cr\,{\sc ii}} 
lines were employed.
We used up to 21 neutral \mbox{Ni\,{\sc i}} lines for the abundance 
determinations. 
The abundances of Ti, Cr, and Ni appear to be very well determined.

We used the MOOG $blends$ option to derive abundances from
uncontaminated but hyperfine-split transitions of \mbox{Sc\,{\sc ii}},
\mbox{V\,{\sc i}}, and \mbox{Co\,{\sc ii}}.
The mean cluster abundances of Sc and V are almost exactly solar, while Co
appears to have a small underabundance (Table~\ref{tab9}).

Abundances of the remaining Fe-group elements were determined by 
synthetic spectrum matching. 
As discussed in \S\ref{linelists} we analyzed three 
\mbox{Mn\,{\sc i}} lines located around 6016 \AA. 
The mean cluster abundance is somewhat subsolar: All RGs slightly 
deficient in Mn; the abundance mean is
$\langle$[Mn/Fe]$\rangle$ = $-0.12$ ($\sigma$ = 0.04).
The Cu abundance of the cluster, based only on the 5782~\AA\ transition
(\S\ref{linelists}), is only marginally subsolar (Table~\ref{tab9}).
We also derived the Zn abundance using the neutral Zn line located at 
6362.3 \AA. 
This absorption line is contaminated by CN, Sc and V, which makes an 
accurate abundance determination difficult for Zn. 
Star-to-star Zn abundance scatter can be seen in Table~\ref{tab9},
but the abundance mean is only slightly in excess of solar.

\begin{figure}
  \leavevmode
      \epsfxsize=8.6cm
      \epsfbox{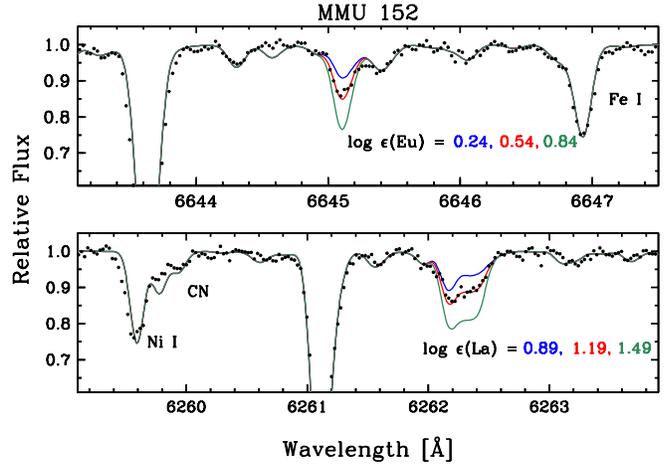}
       \caption{Observed spectra (filled circles) and 
                synthetic spectra (lines) of the \mbox{Eu\,{\sc ii}} 
                6645~\AA\ line (top panel) and the \mbox{La\,{\sc ii}}
                6262~\AA\ lines in MMU152.
                Assumed abundances for the synthetic spectra are
                written in the figure panel legends.
                The synthesis that best matches the observed spectrum
                is the red line; syntheses for abundances smaller
                and larger than this value by 0.3~dex are displayed
                with blue and green lines, respectively.}
     \label{fig7}
\end{figure}

\textbf{\textit{Neutron-capture elements}}:
Abundances of all n-capture elements were determined by spectrum synthesis.
We derived abundances of s-process (slow neutron-capture) elements Y, La 
and Nd, and also one r-process (rapid neutron-capture) element Eu. 
For Y and La we derived solar abundances (Table~\ref{tab9})
from our spectrum analyses of the lines listed in Table~\ref{tab4}. 
Nd abundances were derived from three Nd lines (5249.6, 5293.2 and 5319.8 \AA) 
in most cases, leading to the largest departure from solar abundances of
elements that do not participate in H-burning: 
$<$[Nd/Fe]$>$ = $+$0.18 ($\sigma$ = 0.04).
The lines of the r-process element Eu used for the abundance analysis 
are \mbox{Eu\,{\sc ii}} 6645.1 and 7217.5 \AA.
For some RGs we could not use 7217.5 \AA\ line since its strength is small and 
disappears in the continuum. 
An example for synthetic-observed spectra comparison is given 
in Figure~\ref{fig7} for the \mbox{Eu\,{\sc ii}} 6645.1 and 
\mbox{La\,{\sc ii}} 6262.3 \AA\ regions of MMU~152.
The figure contains three synthesis with 0.3 dex steps and abundances are in 
log~$\epsilon$ unit.
Red solid line in the middle represents the best fit and the abundance accepted for this star.
The individual Eu abundances among all listed RGs 
are in agreement and the mean Eu abundances for the cluster is solar
(Table~\ref{tab9}).

Recent study of \cite{bla} investigates the chemical abundances
of 14 elements (Na, Mg, Si, Ca, Sc, Ti, V, Cr, Mn, Fe, Co, Ni, Y, Ba)
for one RG member of NGC 6940. We have 13 elements that are common
with the sample of \citeauthor{bla}.
All are in good agreement with their findings except for the abundances of
[Na/Fe], [Mn/Fe] and [Co/Fe], which are away from our findings by  0.14 dex, 0.14 dex and $-$0.13 dex, respectively. 

\subsection{Light elements}\label{proton}

LiCNO abundances were derived by spectrum synthesis. 
C, N, and O are bound together through the molecules 
such as CN, CH and CO. 
Therefore, we performed iterative syntheses to derive their abundances. 
As discussed in Paper~I, we used recent molecular laboratory data  
in analyses of the C$_2$ and CN bands.
First we derived O abundances using the non-LTE free [\mbox{O\,{\sc i}}] 
6300.3 \AA\ forbidden line. 
Then C abundances were determined from C$_{2}$ Swan bandheads 
at 5160 \AA\ and 5631 \AA, adopting the new O abundances.  
These steps were repeated until the changes in O and C abundances were 
negligible between iterations.
Finally, with these abundances we analyzed the 7995-8040 \AA\ region 
using $^{12}$CN and $^{13}$CN red system lines to derive N abundances 
and carbon isotopic ratios.

\begin{figure}
  \leavevmode
      \epsfxsize=8.2cm
      \epsfbox{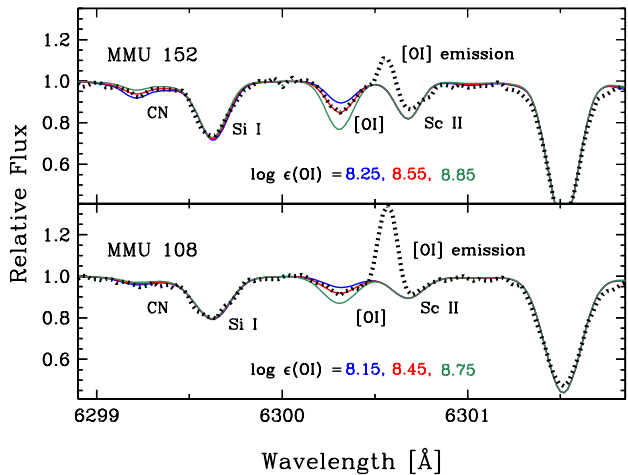}
       \caption{Observed and synthetic spectra of
                the [\mbox{O\,{\sc i}}] line in MMU 152 (top panel)
                and MMU 108 (bottom panel).
                The symbols and lines have the same meanings as in
                Figure~\ref{fig7}.}
     \label{fig8}
\end{figure}

A concern for the O analysis is the relatively small
radial velocity shift of NGC~6940 (Table~\ref{tab3}), which potentially
can lead to contamination of the stellar [\mbox{O\,{\sc i}}] 6300~\AA\ line
by its night-sky emission counterpart.
We also tried to derive O abundances from the [\mbox{O\,{\sc i}}] 6363 \AA\  
without much success: this line is very weak, contaminated by stellar
CN features, and very much distorted by its night sky emission.
Figure~\ref{fig8} shows the best (upper panel) and worst (lower panel) case 
scenario for the 6300~\AA\ situation. 
We decided not to attempt removal of the night sky components in order not 
to risk damaging the [\mbox{O\,{\sc i}}] absorption lines. 
In Figure~\ref{fig8} we compare the observed spectra of MMU~152 and MMU~108 
with synthetic spectra. 
The synthetic spectrum represented with red solid line is the best fit. 
Blue and green solid lines represent the analysis $\pm${0.3} dex away from 
the best fit. 
All of our programme RGs are slightly or moderately affected by the night 
sky emission line.
Since the night sky emission line is more blended with 6300.68 \AA\ 
[\mbox{Sc\,{\sc ii}}] line, we mainly focused on the blue side of the 
[\mbox{O\,{\sc i}}] to obtain more reliable O abundances.
We also took into account the CN 6300.27 \AA\ and \mbox{Ni\,{\sc i}} 
6300.34 \AA\  \citep{joha}
contributions to the [\mbox{O\,{\sc i}}] 6300.31 \AA\ line.

The derived O abundances of NGC~6940 RGs indicate some 
star-to-star scatter, from [O/Fe] = $-0.27$ (MMU~132) to $-0.06$ (MMU~138).
Clearly the reader should keep in mind that the night sky emission 
contamination may increase the uncertainty in our O abundances.
The mean for the cluster is [\mbox{O\,{\sc i}}/Fe] = $-$0.15 ($\sigma$ = 0.06). 
Note that similar O underabundances were also observed among NGC~752 
members (Paper~1), and that cluster's radial velocity safely distances
the stellar and night-sky 6300~\AA\ lines.

\begin{figure}
  \leavevmode
      \epsfxsize=8.2cm
      \epsfbox{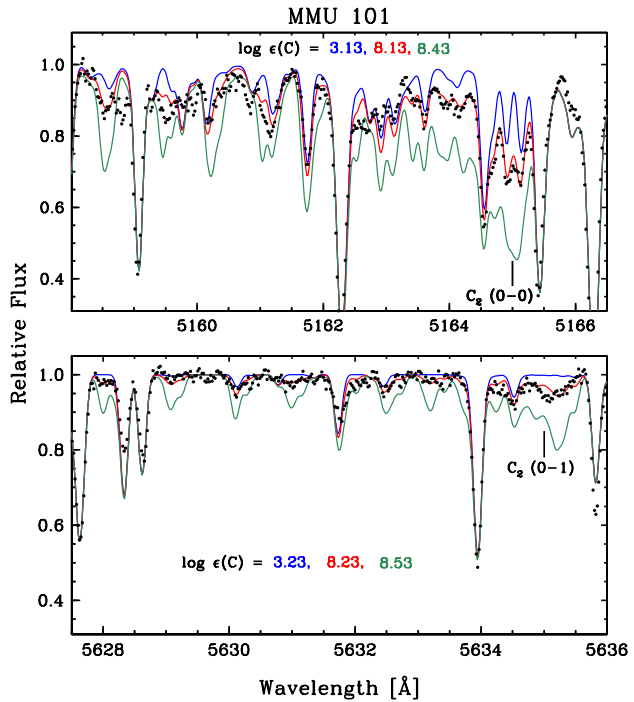}
       \caption{Observed and synthetic spectra of
                the C$_2$ Swan system (0$-$0) bandhead (top panel)
                and (0$-$1) bandhead (bottom panel) in MMU 101.
                The symbols and lines have the same meanings as in
                Figure~\ref{fig7}.}
     \label{fig9}
\end{figure}

Carbon abundances were derived from C$_{2}$ Swan system
features in two regions: the (0-0) bandhead at 5165 \AA\ and (0-1) bandhead 
at 5635 \AA.
Our spectral coverage does not include CH G-band 4300~\AA\ region, so
we were not able to use CH to check our C$_{2}$ results.
The complex structure of the C$_{2}$ bands, which are severely blended 
atomic absorption features in the 5165~\AA\ region and are very weak in
the 5635~\AA\ region, bring difficulties in determination of C abundances. 
In Figure~\ref{fig9}, we show an example spectral synthesis of both 
Swan regions for MMU~101. 
The red solid line represents the best fit to the observed spectrum, and blue 
and green solid lines represent the $-$5 and $+$0.3 dex departure from the best fit 
respectively.
C abundances obtained from 5165 \AA\ bandhead 
are lower compared to those
obtained from 5635 \AA, the mean difference being $\sim0.17$ dex. 
The averages of these two regions were adopted as the final carbon abundances,
which are given in Table~\ref{tab9}.  
All C abundances are subsolar, ranging from [C/Fe]$=-0.49$ (MMU~152) to
$-0.22$ (MMU~138), leading to the mean $\langle$[C/Fe]$\rangle$ = $-0.32$, 
with only modest star-to-star scatter ($\sigma$ = 0.07).

\begin{figure}
  \leavevmode
      \epsfxsize=8.5cm
      \epsfbox{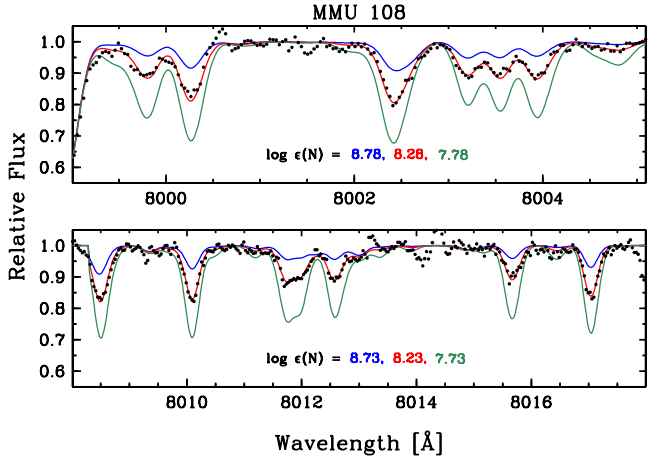}
       \caption{Observed and synthetic spectra of CN 
                transitions in 7999$-$8005 \AA\ (top panel)
                and the 8008$-$8017 \AA\ (bottom panel) wavelength
                regions in MMU 108.
                The symbols and lines have the same meanings as in
                Figure~\ref{fig7}.}
     \label{fig10}
\end{figure}

Nitrogen abundances were derived using $^{12}$CN features in the 
7995$-$8040 \AA\ region.  
All of the RG members in our list are overabundant in nitrogen.
Figure~\ref{fig10} shows the spectrum synthesis for the most N-enhanced 
NGC~6940 member MMU~108, which has [N/Fe] = $+$0.37.
The best fit to the observed spectrum (black dots) is represented by a
red solid line in both panels. 
The average N abundance for this star is log $\epsilon$(N) = 8.26. 
The mean N abundance for the cluster is $\langle$[N/Fe]$\rangle$ = $+$0.23 
($\sigma$ = 0.06).

\begin{figure}
  \leavevmode
      \epsfxsize=8.5cm
      \epsfbox{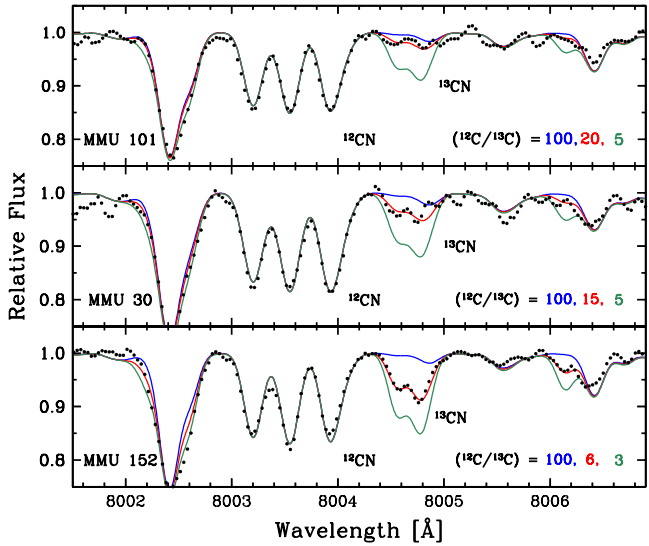}
       \caption{Observed and synthetic spectra of CN  
                transitions in 8001$-$8007 \AA\ wavelength region in
                three stars with different carbon isotopic ratios.
                This spectral interval is shown to highlight the
                prominent $^{13}$CN 8004~\AA\ triplet, which is often
                the key \ciso\ ratio indicator in RGs.
                The symbols and lines have the same meanings as in
                Figure~\ref{fig7}.}
     \label{fig11}
\end{figure}

One of our main goals is to derive \ciso\ ratios, which 
are sensitive indicators of stellar interior light element synthesis 
and envelope mixing in RG stars.
There are several $^{12}$CN and $^{13}$CN features that 
can be used to derive \ciso\ in the $\sim$8000-8048 \AA\ region.
However, due to atomic and molecular line blending, we 
concentrated on analyses of the prominent $^{12}$CN lines at 8003 \AA\ and 
$^{13}$CN lines at 8004 \AA\, and used the strengths of weaker $^{13}$CN 
lines as corroborating evidence. 
In Figure~\ref{fig11}, we display the spectral region surrounding these
CN lines in three different RGs. 
These stars are representative of the \ciso\ range of our NGC~6940 RGs;
see Table~\ref{tab11}. 
As seen in Table~\ref{tab11}, seven cluster members have \ciso\ values in
the 20$-$25 range, as predicted from standard first dredge-up (FDU) theory
for RGB stars.
However, four NGC~6940 program stars have values lower than expected, 
10$-$15.
The most interesting RG is MMU~152 (see the bottom panel of 
Figure~\ref{fig11}), which has \ciso~$=6$.  

\begin{table}
  \centering
\begin{minipage}{85mm}
  \caption{Carbon isotopic ratios.
  \label{tab11}}
  \centering
  \begin{tabular}{@{}lcc@{}}
  \hline
Star  &  $^{12}C/^{13}C$  &   $\sigma$ \\
\hline
MMU~28	&	20 &	$-5 ~/+5$	\\
MMU~30	&	15 &	$-3 ~/+3$	\\
MMU~60	&	20 &	$-3 ~/+5$	\\
MMU~69	&	10 &	$-5 ~/+5$	\\
MMU~87	&	25 &    $-5 ~/+5$	\\
MMU~101	&	20 &	$-5 ~/+5$	\\
MMU~105	&	15 &	$-2 ~/+3$	\\
MMU~108	&	20 &	$-3 ~/+5$	\\
MMU~132	&	20 &	$-5 ~/+5$	\\
MMU~138	&	12 &	$-2 ~/+3$	\\
MMU~139	&	13 &	$-3 ~/+2$	\\
MMU~152	&	6  &	$-1 ~/+2$	\\
$\delta$~Tau	   &	25	&	$-2 ~/+2$\\
$\varepsilon$~Tau  &	26	&	$-1 ~/+4$\\
$\gamma$~Tau	   &	25	&	$-2 ~/+2$\\
$\theta$~Tau	   &	27	&	$-2 ~/+3$\\
\hline
\end{tabular}
\end{minipage}
\end{table}

\begin{table*}
 \centering
 \begin{minipage}{155mm}
  \caption{Updated NGC~752 abundances.}
  \label{tab12}
  \begin{tabular}{@{}lrrrrrrrrrrr@{}}
  \hline
  Species  & \multicolumn{11}{c}{MMU} \\
  & 1 & 3 & 11 & 24 & 27 & 77 & 137 & 295 & 311 & 1367 & 	$<$[X/Fe]$>$ \\
 \hline
 $[\rm{N/Fe}]$\footnote{Only solar abundance updated.} 	&	$	0.29	$	&	$	0.22	$	&	$	0.24	$	&	$	0.28	$	&	$	0.32	$	&	$	0.27	$	&	$	0.28	$	&	$	0.28	$	&	$	0.27	$	&
	$	0.25	$	&	$	0.27	\pm	0.03	$	\\
$[${\mbox{Mn\,{\sc i}}/Fe}$]$	&	$	-0.18	$	&	$	-0.19	$	&	$	-0.21	$	&	$	-0.18	$	&	$	-0.15	$	&	$	-0.15	$	&	$	-0.18	$	&	$	-0.21	$	&	$	-0.21	$	&	$	-0.16	$	&	$	-0.18	\pm	0.03	$	\\
$[${\mbox{Cu\,{\sc i}}/Fe}$]$	&	$	-0.14	$	&	$	-0.19	$	&	$	-0.17	$	&	$	-0.14	$	&	$	-0.05	$	&	$	-0.10	$	&	$	-0.07	$	&	$	-0.08	$	&	$	-0.16	$	&	$	-0.15	$	&	$	-0.12	\pm	0.05	$	\\
$[${\mbox{Y\,{\sc ii}}/Fe}$]$	&	$	-0.04	$	&	$	-0.11	$	&	$	-0.10	$	&	$	-0.08	$	&	$	-0.04	$	&	$	-0.02	$	&	$	-0.03	$	&	$	-0.06	$	&	$	-0.13	$	&	$	-0.16	$	&	$	-0.08	\pm	0.05	$	\\
$[${\mbox{Nd\,{\sc ii}}/Fe}$]$	&	$	0.20	$	&	$	0.15	$	&	$	0.15	$	&	$	0.08	$	&	$	0.22	$	&	$	0.28	$	&	$	0.23	$	&	$	0.22	$	&	$	0.17	$	&	$	0.12	$	&	$	0.18	\pm	0.06	$	\\
\hline
\end{tabular}
\end{minipage}
\end{table*}

\begin{figure}
  \leavevmode
      \epsfxsize=8.5cm
      \epsfbox{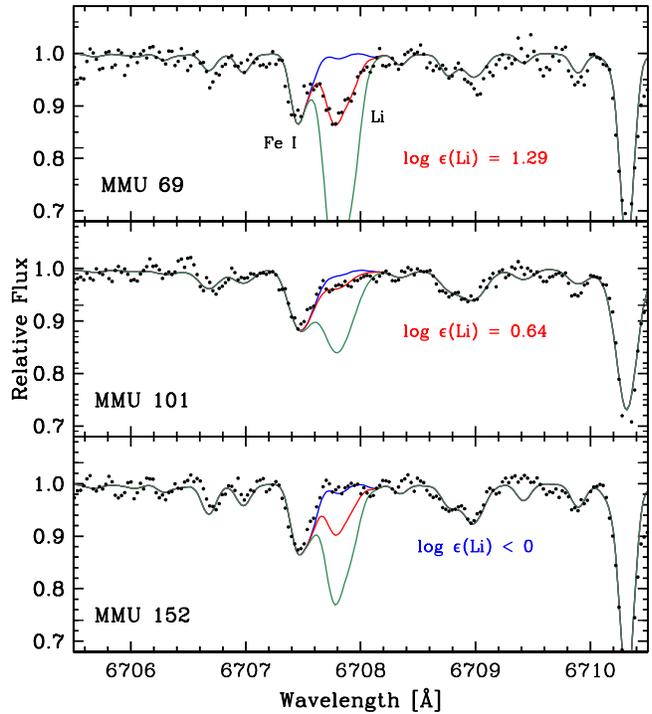}
       \caption{Observed and synthetic spectra of the
                \mbox{Fe\,{\sc i}} 6707~AA\ resonance doublet in
                three stars with contrasting Li abundances.
                The symbols and lines have the same meanings as in
                Figure~\ref{fig7}.}
        
     \label{fig12}
\end{figure}

Li abundances were derived from 
the neutral Li resonance doublet at 6707.8 \AA.  
Our syntheses took into account its hyperfine structure, and also the 
blending with 6707.4 \AA\  \mbox{Fe\,{\sc i}} line.  
In Figure~\ref{fig12} we have plotted the Li region for three RG members.  
These stars represent the member-to-member Li abundance variation in our 
NGC~6940 sample.  
Li abundances of all the RGs are listed in Table~\ref{tab9}. 
MMU~69 has the most enhanced Li abundance among all 
of our targets: log $\epsilon$(Li)~=~1.29.
There are no Li detections for four of the RG members.

\subsection{Abundances in the Hyades}

Derivation of atmospheric parameters and abundances of 
the four Hyades RGs have been described in \S\ref{models} and 
\S\ref{abunds}.
These were accomplished with the same methods that were applied to 
the NGC~6940 RGs, and the results are entered in Tables~\ref{tab6}, 
\ref{tab10} and \ref{tab11}.

The Hyades metallicity determined by our methods is
$\langle$[\mbox{Fe\,{\sc i}}/H],[\mbox{Fe\,{\sc ii}}/H]$\rangle$ = $+0.12$, 
slightly larger than our value for NGC~6940, 
$\langle$[\mbox{Fe\,{\sc i}}/H],[\mbox{Fe\,{\sc ii}}/H]$\rangle$ = $+0.04$. 
This is in good accord with many previous studies that have found
the Hyades to be more metal-rich than the Sun.
For example, $\langle$[Fe/H]$\rangle$ = $+0.16$, $\sigma$ = 0.02 \citep{liu16},
$+0.15$, $\sigma$ = 0.06 (\citealt{med13}, unweighted mean),
$+0.21$, $\sigma$ = 0.07 \citep{smi12},
$+0.22$, $\sigma$ = 0.05 \citep{sch06},
$+0.12$, $\sigma$ = 0.02 \citep{care},
$+0.13$, $\sigma$ = 0.05 \citep{pau03}, and earlier studies that are 
cited in these papers.

\begin{figure}
  \leavevmode
      \epsfxsize=8.5cm
      \epsfbox{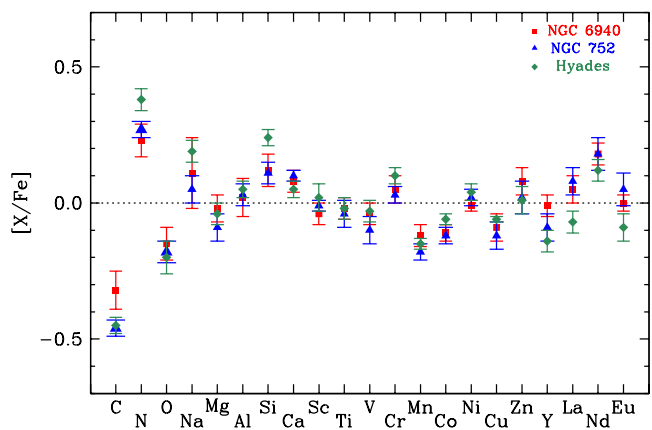}
       \caption{Abundance ratios in the three OCs studied 
                in this work and in Paer~I.
                The point types and colors are identified in the figure
                legend.}
     \label{fig13}
\end{figure}

The relative [X/Fe] values of Hyades giants determined 
in this study are nearly identical to those of NGC~6940 and NGC~752.
Computing the mean difference of the entries in Tables~\ref{tab9}, 
\ref{tab10}, \ref{tab12} and Table 10 in Paper~1, we find 
$\langle$[X/Fe]$_{NGC6940}$ $-$ [X/Fe]$_{Hyades}\rangle$ = 
0.00~$\pm$~0.02 ($\sigma$~= 0.08).
Doing a similar comparison of the Hyades to NGC~752 (Paper~I, Table~10) yields
a similar agreement: $\langle$[X/Fe]$_{NGC752}$ $-$ [X/Fe]$_{Hyades}\rangle$ =
$-0.02$~$\pm$~0.02 ($\sigma$~= 0.08).
This general abundance ratio accord among the three clusters is apparent
in the [X/Fe] values displayed in Figure~\ref{fig13}.

There are few comprehensive chemical composition studies
of Hyades giants in the literature.
Of recent studies, \cite{care} includes the most elements in common with
our work.
In general our Hyades abundances are in good accord with theirs:
$\langle$[X/Fe]$_{Carrerra}$ $-$ [X/Fe]$_{us}\rangle$ =  
$-0.01$~$\pm$~0.02 ($\sigma$~= 0.08, for 16 species in common).
\cite{lam81}, one of the first
large-sample surveys of light elements in RGs, included the four Hyades stars.
Forming abundance differences for individual stars in the sense 
\citeauthor{lam81} $minus$ this study, and computing 4-star means, we find 
$\Delta$[Fe/H] = $-0.02$ ($\sigma$ = 0.04),
$\Delta$[C/Fe] = $+0.16$ ($\sigma$ = 0.04),
$\Delta$[N/Fe] = $+0.01$ ($\sigma$ = 0.03), and
$\Delta$[O/Fe] = $+0.15$ ($\sigma$ = 0.04).
Caution is warranted in this comparison, given the improvements in stellar
atmosphere modeling and atomic/molecular linelists over the last few decades,
and with our more extensive spectral coverage of the C$_2$ Swan bands 
compared to the ealier study.
We suspect that the larger O abundances of \citeauthor{lam81} may be due to
the influence of the \mbox{Ni\,{\sc i}} contamination of the 6300 \AA\
[\mbox{O\,{\sc i}}], which was not understood at the time of the
\citeauthor{lam81} study.
Their C abundances are somewhat larger than ours, but 
their N abundances are in good accord.
\citeauthor{lam81} quote carbon isotopic ratios from earlier papers in 
their series (e.g. \citealt{tom}), which are on average somewhat lower 
than ours: $\Delta$\ciso\ = $-5$ ($\sigma$ = 2).
Their values should generally be more accurate than ours since they
used more CN features in their work than done here.
More detailed comparisons are beyond the scope of our work here.

\section{SUMMARY AND DISCUSSION}\label{discuss}

This study is the second in this series of papers on the
chemical compositions RGs of OCs.
Here we report the first detailed chemical abundance analysis of NGC~6940, 
giving results for 12 RG members.
In Paper~1 we studied 10 RG members of the open cluster NGC~752 in 
the same manner. 
Our main focus for these two studies is to derive the abundances of 
light (LiCNO) elements and \ciso\ ratios.

We first confirmed NGC~6940 membership of our 12 program stars
via RV calculations from our spectra.
By applying the Y$^{2} $ isochrones to \cite{lar60} photographic observations 
we then estimated the turn-off mass ($\simeq$2 M$_{\odot}$) and 
age ($\sim1.1$~Gyr) of the cluster. 
The turn-off mass was assumed to apply to the RG masses.
Then with the known cluster reddening, distance modulus, 
colors and spectroscopic line depth ratios, we calculated
initial model atmosphere parameters. 
We iteratively applied standard spectroscopic line 
analysis techniques to iterate to achieve final model atmosphere parameters
\teff, \logg, \vmicro\ and [M/H].
The mean metallicity for the NGC 6940 was 
found to be about solar, $\langle$[M/H]$\rangle$ = $-$0.06 $\pm$ 0.07.
 
We then derived the abundances of a number of light 
(Li, C, N, O), odd-Z (Na, Al), $\alpha$ (Mg, Si, Ca), Fe-group 
(Ti, Cr, Ni, Mn, Cu, Zn), n-capture elements (Y, La, Nd, Eu) and
also determined the \ciso\ isotopic ratios of the RGs. 
The overall abundance pattern of the NGC~6940 members are mostly 
around solar (Figure~\ref{tab6}).
For some of the species, such as Na, line-to-line scatter among 
individual RGs is a bit higher compared to others. This is either due to lack 
of recent lab data or complex hyperfine structures of these species. 
Along with the RG members of NGC~6940 (1.1 Gyr), 
we also analyzed the four well-known RG members of the Hyades (0.63 Gyr) 
open cluster with the aim of standardizing and showing the reliability of the 
method we use for the analysis of the open cluster NGC~6940. Hyades RGs 
have been a subject of many studies. 
Our derived atmospheric parameters and elemental 
abundances of Hyades RGs as listed in Table~\ref{tab6} and 
Table~\ref{tab10}, respectively.
Our results are generally in good agreement with the literature.
We also give the overall comparison
of the abundances in NGC~6940, NGC~752 and Hyades open clusters 
in Figure~\ref{fig13}. Some deficiency in neutron-capture elements Y, La and 
Eu and a slight enhancement in Si abundance are noticeable in 
Hyades open cluster when compare to NGC~6940 and NGC~752. 
Other elements mostly behave similarly.

\subsection{Evolutionary Status of the RGs}

NGC~6940 is an intermediate age open cluster with an 
age of 1.1 Gyr suggested by Y$^{2}$ isochrone fits to its CMD 
(Figure~\ref{fig1}). 
The LiCNO abundances and \ciso\ ratios of the RGs we investigate in this
study indicate that although these stars are the members of the same OC, 
they must have experienced different types/amounts of mixing during their 
giant branch evolution.
In our sample, we have six RGs that have \ciso\ $<$ 20, lower than predicted by 
canonical models: MMU~30, 69, 105, 138, 139 and 152 (Table~\ref{tab11}).
Four of the RGs have no sign of Li in their spectra: 
MMU~30, 105, 108 and 152 with \ciso = 15, 15, 20 and 6, respectively. 
In the next two subsections we will discuss the general
features of Li, \ciso, [N/C] and Na abundances in NGC~6940, but we call
attention here the results for two of our stars.
MMU~152 stands out in a need for an extra-mixing process. 
MMU~108 has the highest \teff\ among other RGs with no sign of Li and 
\ciso\ $=$ 20.  
Similarly, other members with low \ciso\ ratios might have experienced 
extra-mixing processes at different levels and started 
He-burning in their cores. 
As first suggested by \cite{can}, a clump of red giants centred 
near $M_{V_0}$ = +1 and $(B-V)_{0}$ = 1 in the general field are 
probably core helium burning horizontal branch stars. 
The mean absolute magnitude and intrinsic colour for the RG members of 
NGC~6940 in our sample are $<$$M_{V_0}$$>$ = 1.12 ($\sigma$ = 0.32) and 
$<$$(B-V)_{0}$$>$ = 0.90 ($\sigma$ = 0.06), 
suggesting all the RGs investigated here reside at the clump region. 
Most of the sample seem to locate closer to the bluer part of the
RC, which is generally called the red horizontal branch (RHB) 
(e.g. \citealt{kaem}),
indicating the slight initial mass differences between the members.

\subsection{LiCNO}

\begin{figure}
  \leavevmode
      \epsfxsize=8.5cm
      \epsfbox{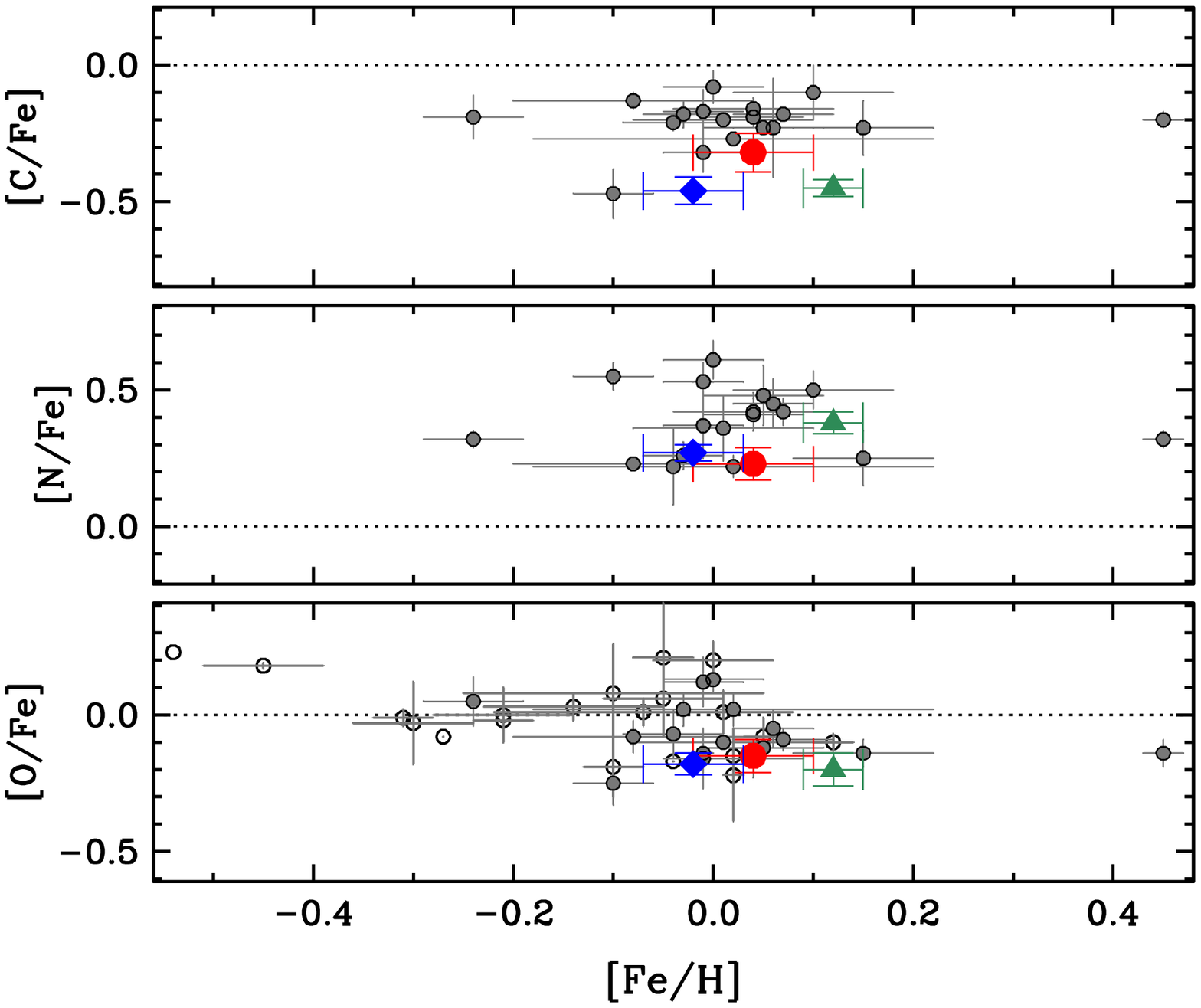}
       \caption{Comparison of CNO abundances with the literature. Red dots are   
       NGC~6940, blue
       diamonds are NGC~752 while greed triangles are Hyades. 
       Grey dots \citep{smi,tau00,tau05,tau15,mik10,mik11,mika,mik12,san13,zacs}
       represents OCs from the literature and open circles 
       \citep{friel05,friel10,jacob08,jacob09,jacob11,pan10,yon05}
       are O abundances derived without C and N abundances. }
     \label{fig14}
\end{figure}

The so-called FDU mechanism alters
surface abundances as stars evolve up the RGB.
During this episode, the convective envelope of a star deepens towards 
inner regions and brings the chemically processed material to the surface.
This mixing causes surface abundance changes in RGs:
typically [N/Fe] increases by about 0.4 dex, [C/Fe] decreases by about 
0.20 dex, and O abundance remains 
around solar (e.g. \S2.2 in \citealt{kar14}).
The \ciso\ isotopic ratio also changes during the FDU, 
dropping from its main sequence value ($\sim$90) to 20$-$30. 
The He abundance increases by $\Delta$Y~$\approx$~0.012 (unobservable).
The main isotope of Li, $^{7}$Li, is destroyed by H burning at relatively low 
temperatures (T $\geq$ 2.5 $\times$ 10$^{6}$~K). 
Therefore, its surface abundance also alters during the FDU
and drops to log ($\varepsilon$)(Li)~$\simeq$~1 
\citep{kar14}.
Besides the elements mentioned above, FDU has minor effects on the 
other elements \citep[see][and references therein]{kar14}.

Li is a fragile element with short lifetime in 
H-burning zones. 
It surface abundance is known to be very sensitive to mixing processes, 
thus can vary greatly among RG stars.
The complex behaviour of Li has been studied by many authors for decades.
For example, \cite{palme} investigated the Li abundance problem 
in low-mass ($\leq$ 3 M$_{\odot}$) solar metallicity red giants. 
They applied several theoretical models that involve
deep-mixing mechanisms to the observational data. 
They concluded that different set of deep-mixing mechanisms can cause 
different amount of Li abundances lower than A(Li) = 1.5 after the FDU. 
Their models do not predict any further mixing after the luminosity function 
bump, during which
H-burning shell erases the chemical discontinuity left by the FDU. 
Results of the mixing at different depths are shown in Figures~1 and 4
of \cite{palme}.
Our sample has a Li abundance range from
0 $<$ A(Li) $<$ 1.3. 
Assuming that the theoretical timescales for first-ascent giants are in
good agreement with the observations, the
 average temperature of the RGs in our sample along with the turn-off mass 
of the NGC 6940 strongly suggest that our RGs reside at the clump region 
predicted in \cite{palme}.   
In another recent study,  \cite{char} discuss
how the thermohaline and rotation-induced mixing modify the surface
abundances in giant stars. 
Their model predictions indicate that, for a star 
with initial mass of 2 M$_{\odot}$, depending on the mixing 
mechanisms (thermohaline instability and/or rotation-induced mixing) 
involved at the end of the FDU, Li abundances can change 
from about the canonical value of 1.4 to $<$0 (Figure 14 in  \citealt{char}). 
The Li abundance diversity in our sample appears to be a very good
implication of the different mixing mechanisms taking place in stellar 
interiors. 

The mean CNO abundances we obtained for the NGC~6940 are 
[C/Fe] = $-$0.32 $\pm$ 0.07, [N/Fe] = $+$0.23 $\pm$ 0.06, 
[O/Fe] = $-$0.15 $\pm$ 0.06, which are in general agreement with the 
standard mixing theories summarized above.
In Figure~\ref{fig14}, we compare our CNO abundances with  
other RGs in different OCs studied in the literature.
In this figure grey dots represent the average abundance ratios of those RGs.
In the bottom panel of the Figure~\ref{fig14} open circles represent
the O abundances derived without C and N abundances. 
The average CNO abundance ratios of the NGC~6940, NGC~752 and Hyades are 
illustrated by red dot, blue diamond and green triangle, respectively. 
The error bars indicate the standard deviation of the mean abundances. 
As is seen in this figure,
our CNO abundances agree fairly well with the data from other open clusters. 
Emphasis on the usage of different solar abundances in these studies 
should be made. 
For example, solar CNO abundances we used in our study (Table~\ref{tab8}) 
are 8.43, 7.99 and 8.69, respectively. 
Other authors, however, usually prefer to adopt the solar abundances from 
\cite{gre98}, in which the solar CNO abundances are given as 8.52, 7.92, 8.83,
or from \cite{gre07} that reports abundances as 8.39, 7.78 and 8.66, 
respectively. 
These differences, in general, create small offsets 
that affect the comparisons of abundance ratios.

Theoretical models predict that post dredge-up [N/C] 
values increase with initial stellar masses (\citealt{char}). 
Derived [N/C] values in our clusters are moderately in accord with the 
predictions of \cite{char}.
Average [N/C] values of the clusters are 
[N/C]$_{\textrm{NGC~6940}}$ = 0.56 $\pm$ 0.11, 
[N/C]$_{\textrm{NGC~752}}$ = 0.73 $\pm$ 0.04
and [N/C]$_{\textrm{Hyades}}$ = 0.83 $\pm$ 0.02. 
Since our C and N abundances were determined in the 
same way for these three clusters, their relative N/C ratios may be
more robust that derived values for individual stars in individual clusters.
We may attempt to interpret these relative abundances
in terms of theoretical predictions discussed in \cite{char} and \cite{kar14}.

[N/C] ratios among the giant members of NGC~6940 have a significant range
from 0.44 to 0.83. 
MMU~30 has the lowest [N/C] value with 0.44 and has 
\ciso\ $\simeq$ 15. 
Evaluating these parameters together with the turn-off mass of the
cluster in Figure 17, 19 and 20 of \cite{char}
suggests that thermohaline 
mixing is the major mechanism that leads to these values, perhaps along 
with a slightly high initial rotational velocity. 
On the other hand, MMU~152 with [N/C] = 0.83 and \ciso\ $\simeq$ 6 
exhibits a unique behaviour. Comparing its values with the model 
predictions indicates a mixing mechanism mainly governed by a 
very high
initial rotational velocity (V$_{\textrm{ZAMS}}$ $>$ 300 \kmsec) that 
might also imply a higher initial mass. 
Diversity in [N/C], \ciso~ratios and Na abundances 
(discussed below) suggests that, although the cluster 
members have started of similar initial chemical compositions, perhaps
different initial masses distribution of the members prompted different
mixing mechanisms.

The mean [N/C] = 0.73 of NGC~752 suggests a contribution
by rotation-induced mixing. 
Different \ciso\ ratios among the cluster members (Paper~1) imply the 
existence of a dominant thermohaline mixing mechanism in some cases. 
Evolved members of Hyades have more depletion in carbon (previously 
noticed by \citealt{sch09}) compared to NGC~752 and NGC~6940.
The average [N/C] = 0.83 value of Hyades suggests higher initial masses and 
rotation-induced mixing with relatively high initial velocities. 
\ciso\ values, on the other hand, are in accord with the standard models. 
Our interpretation on the mechanisms that lead to extra-mixing in RGs are based 
on various parameters and need to be taken with caution.

\begin{figure}
  \leavevmode
      \epsfxsize=8.5cm
      \epsfbox{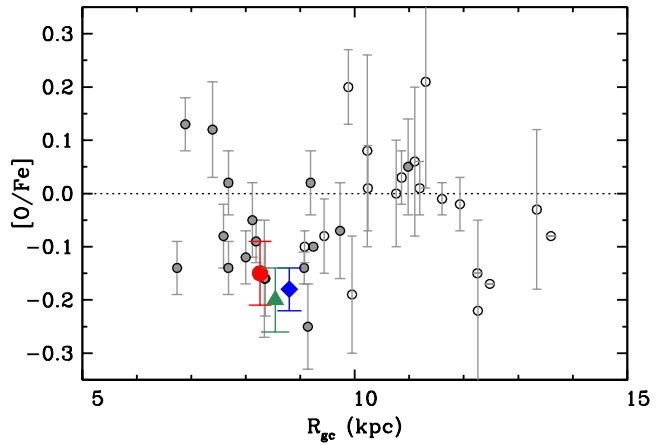}
       \caption{Galactocentric radial distribution 
                           of [O/Fe] abundances. 
                           Symbols are the same as in Figure~\ref{fig13}.}
     \label{fig15}
\end{figure}

The elemental abundances we derived for the RGs of Hyades 
agree well with the literature (e.g. \citealt{care}) but some discrepancies 
arise for a few elements. 
For example, oxygen is generally found around solar in most cases but our 
sub-solar oxygen abundances are only in accord with \cite{care}. 
Subsolar oxygen abundances seem to be a common issue among
the giant members of NGC~6940, NGC~752 and Hyades (Figure~\ref{fig13} and 
Figure~\ref{fig15}). 
On average, the oxygen abundance in these clusters is subsolar about 0.2 dex.
A simple explanation to this deficiency is that these clusters, 
which have galactocentric 
distances between 8.2 $<$ $R_\textrm{GC}$ $<$ 8.8 kpc, might have formed from 
molecular clouds with initially low oxygen abundances. 
In Figure~\ref{fig15} we plot O abundance as a function
of Galactocentic distance for the same clusters that were shown in 
Figure~\ref{fig14}.
Our sub-solar [O/Fe] values are shared by clusters at similar 
Galactocentric distances that have been reported in other studies. 
Clusters with $R_\textrm{GC}$ $>$ 10 kpc, and possibly those with 
$R_\textrm{GC}$ $<$ 7.5 kpc, exhibit solar [O/Fe] ratios.

Contribution of OCs to the Galactocentric radial abundance gradient has 
been studied by many authors for several decades. 
First \cite{twa} showed that there is a sharp discontinuity of the [Fe/H] 
distribution at $R_\textrm{GC}$ = 10 kpc, 
which appeared to be the transition region between the
inner and outer Galactic disc. 
Later studies, such as \cite{yon05} and \cite{jac09,jac11}, concluded that 
the transition region lies at $R_\textrm{GC}$ $\approx$ 12-14 kpc by 
analyzing the abundances of more distant OCs.   
The behavior of [O/Fe] - $R_\textrm{GC}$ variation is complicated by 
data incompleteness.
There are two more OCs we did not include in Figure~\ref{fig15} at 
16.4 kpc (Be~20) and 22.9 kpc (Be~29) with supersolar O abundances, 
[O/Fe] = $+$0.18 and $+$0.23, respectively \citep{yon05}.
In general the distribution of [O/Fe] with Galactocentric distance is not 
clear, scattering from $-$0.25 to $+$0.20, and more results for clusters
far from the solar circle are needed.

Finally, \cite{lebz} investigated the oxygen isotopic 
ratios in intermediate-mass giants, estimating $^{16}$O/$^{17}$O and 
$^{16}$O/$^{18}$O ratios in six red giant members of selected relatively
nearby open clusters with known main sequence turnoff masses.
They obtained high-resolution H and K-band spectra of these targets and 
used several OH lines to measure the [$^{16}$O/H] abundances in these giants. 
They found an average subsolar value, [$^{16}$O/H] = $-$0.25 $\pm$ 0.03 
for their sample. 
They concluded that the initial $^{16}$O abundance in the protostellar 
cloud is directly related to the final observed deficiency in oxygen in
RG's, which can only be achieved
with models have initially subsolar $^{16}$O values.

\subsection{Na} 
 
Na abundances show different behaviours in each cluster 
in our sample.  
While RGs of the NGC~6940 and Hyades open clusters indicate Na 
overabundances, NGC~752 members have approximately solar Na abundances 
(Paper~1).
\cite{smi12} determined Na contents in three Hyades giants,
$\delta$~Tau, $\varepsilon$~Tau and $\gamma$~Tau, and found an average 
non-LTE corrected [Na/Fe] = $+$0.3 dex; this is about 0.1 dex higher than 
what we have found in this study. 
The Na lines used for the abundance determination in both studies are the same
(including $gf$ values) except for the 5682.6 \AA\ line, which was 
excluded by \cite{smi12} due to a \mbox{Cr\,{\sc i}} blend. 
Eventually they used only 6154.2 and 6160.8 \AA\ lines to compute the Na 
abundances. 
As reported in \cite{tak03}, the 6154 and 6161 \AA\ lines are insignificantly 
affected by the departures from the LTE ($\leq$ 0.1 dex) and 5683 \AA\ 
line is somewhat affected for metal-poor stars with $-3\leq [Fe/H] \leq-1$. 
By taking into account the contamination from \mbox{Cr\,{\sc i}} blend, 
we found that this line gives Na abundances well in agreement with those 
obtained from 6154.2 and 6160.8 \AA\ lines.
The discrepancy of about 0.1 dex in Na abundance between our and Smiljanic's 
results may be due to slightly different [Fe/H] and 
log $\epsilon$(Na)$_{\sun}$ values in both studies.
Overall, our findings seem to be well in agreement with the non-LTE results 
from \cite{smi12}.

We also observe some enhancement in Na among the RGs of 
NGC~6940; the cluster mean is $\langle$[Na/Fe]$\rangle$ = 0.11 $\pm$ 0.13 dex. 
The high standard deviation is mainly due to MMU~108 and MMU~152, which 
have relatively high Na abundances of $\sim$0.3 dex and $\sim$0.4 dex, 
respectively. 
We also point out the modest Al overabundance in MMU~108 and MMU~152 by 
about 0.14 dex.
These members are also enriched in N by about 0.35 dex, which is above 
the average of the cluster. 
Although overabundances in Na, Al and N abundances mimic the classic
[Na/Fe]-[Al/Fe] and [Na/Fe]-[N/Fe] correlations observed in globular clusters 
(e.g. \citealt{kraft} and references therein), 
this phenomenon is not common among open clusters and
future investigation of a larger sample should be pursued for a definite conclusion
on the distinction of NGC~6940.

Na overabundances in both NGC~6940 and Hyades may be explained by 
evolutionary models. 
\cite{char} investigated the effects of different mixing 
scenarios for solar metallicity stars of low- and intermediate mass 
(1 to 4 M$_{\odot}$).  
They compared theoretical results with observational data from open cluster 
studies and concluded that no change in Na abundance is expected for 
stars with turn-off masses nearly up to 2 M$_{\odot}$, but for higher 
masses one should expect to see Na abundances enhanced by about 0.2 dex 
(see also  \citealt{kar14}). 
In the case of NGC~6940 ($M_\textrm{TO}$$\simeq$2 M$_{\odot}$) and 
Hyades ($M_\textrm{TO}$$\simeq$2.3 M$_{\odot}$), we see Na abundances 
enhanced by about 0.1 and 0.2 dex, respectively.
Comparison of our [Na/Fe] values with turn-off masses (Figure 21 in 
\citealt{char}) of the clusters suggests that different extra-mixing mechanisms 
must have been involved in each cluster. 
According to theoretical models of \citeauthor{char}, 
 younger open cluster Hyades shows consistency with the model predictions 
that include both  thermohaline (th) and
 rotation-induced mixing ($V_\textrm{ZAMS}$);
``th$+$$V_\textrm{ZAMS}$=110 \kmsec" and ``th$+$$V_\textrm{ZAMS}$=250 \kmsec" 
(Figure  21 in \citealt{char}).
Most of the RGs of older NGC~6940, on the other hand, coherent with the standard predictions.
However, with the enhanced Na abundances, MMU~108 and MMU~152 seem
to diverge from the rest of the cluster members, suggesting that these 
two stars share a different evolution history probably with much 
higher initial velocities. 
NGC~752 ($M_\textrm{TO}\simeq1.6 M_{\odot}$, Paper~1)
has an average Na abundance of about solar and matches well with the standard 
theory without the need of any additional mixing mechanisms.

Overall evaluation of the atmospheric parameters, LiCNO and 
Na abundances and \ciso\ ratios of the NGC~6940 RGs along with the theoretical 
predictions discussed above reveals that the stars in our sample are 
probable RC stars and were subjected to different amount/type of 
extra-mixing mechanisms depending on their initial masses.

This is the first extended chemical abundance analysis of the 
open cluster NGC~6940.  
Further investigation of the same members in the infrared H- and K-band 
spectral region using Immersion Grating Infrared Spectrograph 
(IGRINS; \citealt{yuk10}) is underway.

\section*{Acknowledgments}

We thank our referee for very helpful discussions that helped improving our paper.
We also thank Craig Wheeler for helpful comments.
Our work has been supported by The Scientific and Technological 
Research Council of Turkey (T\"{U}B\.{I}TAK, project No. 112T929), by the US
National Science Foundation (NSF, grant AST~12-11585), and by the 
University of Texas Rex G. Baker, Jr. Centennial Research Endowment.
This research has made use of: NASA's Astrophysics Data System 
Bibliographic Services; the SIMBAD database and the VizieR service, 
both operated at CDS, Strasbourg, France; the WEBDA database, operated 
at the Department of Theoretical Physics and Astrophysics of the Masaryk 
University; and the VALD database, operated at Uppsala University, the 
Institute of Astronomy RAS in Moscow, and the University of Vienna.

 \bibliographystyle{mn2e}{}
 \bibliography{gamze}

\end{document}